\documentstyle[preprint,aps,eqsecnum,epsfig]{revtex}

\begin{document}
%\draft
%%
%
\title{Dual vortices and gauge choice in Abelian projected\\
SU(2) lattice gauge theory}
\author{Kenneth Bernstein, Giuseppe Di Cecio and Richard 
W. Haymaker\thanks{bernsten, dicecio and haymaker@rouge.phys.lsu.edu}}
\address{Department of Physics and Astronomy\\
Louisiana State University\\Baton Rouge, Louisiana 70803 USA}
%
%\date{\today}
%
\maketitle
\begin{abstract}

We explore vortex formation for Abelian projected SU(2) in the Polyakov
gauge and compare the results with those calculated in the 
maximal Abelian gauge.  In both gauges, a non-zero vacuum expectation
value of a monopole field operator signals confinement. We find
vortices in the Polyakov projection, confirming 
the connection between the dual superconductor
order parameter and the existence of vortices.   However we find that
the Polyakov Abelian projection is problematic, leaving the maximal
Abelian projection as the best candidate to define an effective theory
of confinement in this scenario.

\end{abstract}

\section{Introduction}

The key principle governing the onset of superconductivity is the
spontaneous breaking of the U(1) gauge symmetry via a 
non-zero vacuum expectation value of a 
charged field \cite{weinberg}.    An immediate
consequence of this is the generation of a photon mass and, for type II
superconductors, the formation of vortices which confine magnetic
flux to narrow tubes
as revealed by the Ginzburg-Landau 
theory\cite{hay,suzuki,gl,abrikosov,tinkham}. 
Lattice studies of dual superconductivity in SU(N) gauge theories seek
to exploit this connection in establishing the underlying
principle governing color confinement.

On a four dimensional lattice, the effective Euclidean
lattice Higgs theory is the appropriate generalization of the
Ginzburg-Landau theory.  The Higgs field is a 0-form living
on the sites and the gauge field is a 1-form living on the links.   
Classical solutions to this 
theory exhibit the connection between the non-zero vacuum expectation
value of the Higgs field and vortex formation.   

In U(1) lattice pure gauge 
theory  (no Higgs field), this same connection is seen to be
present, not in the defining variables,
but rather in the dual variables.   More specifically:
\begin{enumerate}
\item A field with non-zero magnetic monopole  charge, 
$\Phi$,
has been constructed\cite{fm,pw,ppw,giacomo2}. It is a  composite  4-form
living on hypercubes constructed from
gauge fields.  There are also monopole current 3-forms.

On the dual lattice this monopole operator is a 0-form living
on dual sites.  The monopole currents are 1-forms 
living on dual links.  These currents either form closed loops
or end at monopole operators.   

The monopole operator has a non-zero vacuum expectation value 
in the dual superconducting
phase, $\langle \Phi \rangle \neq 0$, 
thereby signaling the spontaneous breaking 
of the U(1) gauge symmetry.
\item  Dual Abrikosov vortices have been seen 
in simulations\cite{shb,hsbw}.  
They are identified
by the signature relationship between the electric field and the
curl of the monopole current in the transverse profile of the
vortex.  The  dual coherence length, $\xi_d$  measures the
characteristic distance from a dual-normal-superconducting
boundary over which the dual-superconductivity  turns on.
The dual London penetration length, $\lambda_d$ measures the attenuation
length of an external field penetrating the dual-superconductor.
 The dual photon mass  $ \sim 1/\lambda_d$ and
the dual Higgs mass $ \sim 1/\xi_d $.
\end{enumerate}
A signal  $\langle \Phi \rangle \neq 0$ without the consequent signal
of a dual photon mass does not imply confinement.  An observation
of a dual photon mass, i.e. vortex formation, 
without $\langle \Phi \rangle \neq 0$ does
not reveal the underlying principle governing the phenomenon.  

The lattice Higgs theory,  treated as an effective theory, i.e. limited 
to classical solutions,  and
considered in the dual sense, provides a model for
interpreting simulations of the pure gauge theory that can reveal these
important connections.

The link of these considerations to confinement 
in  non-Abelian gauge theory  is through the
Abelian projection\cite{thooft1}.  One first fixes the gauge while
preserving U(1) gauge invariance.   The non-Abelian
gauge fields can be parametrized in terms of a U(1) gauge field and charged
coset fields.  The working hypothesis is that operators 
constructed from the  U(1) gauge field alone, i.e. Abelian
plaquettes, Abelian Wilson loops, Abelian Polyakov lines 
and monopole currents,  will exhibit
the correct large distance correlations relevant for confinement. 
But there is as yet no definitive way to choose the gauge which defines 
the Abelian projection and hence no unique way to define
Abelian links and coset fields from the SU(2) links.

We are  seeking
a judicious choice of dynamical variables  --- defined by a particular
Abelian projection ---  which can account for confinement
via an effective theory involving these dynamical variables.  
This should be our first goal.  If this is
solved satisfactorily, then we can investigate how the picture
changes if we go to an alternative set of variables  --- a
different Abelian projection --- since the 
phenomenon we are describing is of course gauge invariant.  We do not
expect that the same mechanism would describe confinement in two
different Abelian projections.  

There is a very nice illustration of this point in a paper by
Chernodub, Polikarpov and Veselov\cite{cpv2}.  They compare
two Abelian projections.
\begin{itemize}
\item The first is the 
maximal Abelian gauge\cite{klsw} which is the most widely studied 
and perhaps the most promising candidate.  Monopoles are the
magnetic charge carriers of the persistent currents.  
\item Second they exhibit an
Abelian projection in which confinement is due to objects other
than monopoles.  They choose the ``minimal Abelian
projection'' and  show that confinement is due to topological objects 
which are denoted ``minopoles''.    
\end{itemize}
Here we have two projections,  two sets of dynamical variables, and two
different descriptions of confinement, both viable candidate mechanisms.

Our goal in this paper is to test the connection between vortex
formation and a non-zero vacuum expectation of the monopole 
field in the same Abelian projection.  There is an intimate
connection between these two results in a Higgs theory and hence 
it is a strong test of the idea for the dual theory. The two
obvious candidate projections are the Polyakov gauge  
and the Maximal Abelian gauge.
\begin{description}
\item[Polyakov Gauge:] Del Debbio, Di Giacomo, Paffuti 
and Pieri\cite{ddpp}  have  constructed  a 
monopole field operator that shows a very strong signal with a 
sharp discontinuity in the vacuum expectation value
 of the monopole operator
 as a function of temperature in the neighborhood of
$T_c$, the deconfining temperature.  Their calculation is manifestly
gauge invariant but a Polyakov line was used in the definition
the monopole operator.   In this sense we associate their calculation
with the Polyakov gauge as we will explain further below.   We
look at vortex formation in the present paper.
\item[Maximal Abelian gauge:] The vortices 
are well established\cite{hsbw,sbh,mes,ph,cc}.  
 The static potential constructed from Abelian links
gives as definitive a signal of confinement
as the gauge invariant static potential as found by 
Suzuki et.al.\cite{suzuki,ss}, Stack et.al.\cite{sw,snw} and
Bali et. al. \cite{bbmps}.  Bali et.al. find the 
Abelian string  tension calculation gives $0.92(4)$ times the 
full string tension for $\beta = 2.5115$.  Whether this approaches
$1.0$ in the continuum limit remains to be seen.

The calculation of Del Debbio et.al\cite{ddpp}
 of the monopole field operator as a 
practical matter is not adaptable to this gauge.   It would
require hundreds of gauge fixing sweeps of the whole lattice in order
to accept or reject a single link update.  
Chernodub, Polikarpov and Veselov\cite{cpv1} have more recently calculated
the constraint effective potential as a function of the monopole
field in this gauge 
and found a symmetry breaking minimum. However they reported
a problem of obtaining statistics and instead calculated an
approximation to the effective potential. 
\end{description}
In this paper we seek  to establish vortex formation 
in the Polyakov gauge. 

More recently Nakamura et al.
\cite{nbekms} studied an alternative monopole operator  
defined in terms of the variables which occur in the monopole 
form of the action.  We have not yet addressed the issue of 
establishing vortex formation in that framework.

We point out some technical difficulties in implementing the
Polyakov Abelian projection.   We use  Abelian Polyakov lines 
to represent static sources.   However they have an anomalous
behavior which we discuss in Sec.IV.  After an appropriate modification
of the sources we find vortices.  However they are suppressed relative
to the analogous calculation in the Maximal Abelian gauge.  Our 
conclusion is that the Polyakov Abelian projection is not likely to 
give a quantitative description of confinement within the the 
confines of this particular scenario.  
The fact that there are vortices in conjunction with a non zero vacuum
expectation value of a monopole operator confirms the connection we
sought.  However that is not enough to get agreement between the
string tensions as this calculation shows.  The 
charged coset fields are thought not to contribute to 
long range physics, but
they are capable of partially screening the sources. 
This consideration does not arise in the U(1)  theory because
there is no charged dynamical field and this effect appears to be 
minimized in the maximal Abelian gauge. 

 We conclude in this paper 
that the screening due to these charged fields is quite large, and  that
is why the vortices are suppressed.  
The signal of this suppression can be seen most easily in measurements of
$ div$$ E$.  Measuring this on  a site coincident with 
the static source shows a very large attenuation
in the Polyakov projection compared to the maximal Abelian projection.
  A recent  measurement of a gauge invariant definition of $ div$$ E$
 by Skala et. al.\cite{sfz} for SU(3) which uses the
Polyakov line to define the adjoint field exhibited no
screening.

To unify the discussion of various 
 choices of the Abelian projection  in Sec. II we stress the
introduction of a composite field $\widehat{\phi}(x)$ constructed 
from the the
gauge configuration $\{U_{\mu}(x)\}$.  
In Sec.~\ref{sec:3} we review the gauge invariant 
definition of field strength in order to connect our calculation
to the monopole operator calculation of Del Debbio et.al.\cite{ddpp} 
In Sec.~\ref{sec:4} we confront problems with the Polyakov gauge
as mentioned above. We must
represent static sources with Polyakov lines in order to study
the theory at finite temperature.  There is an
inherent ambiguity  on the lattice of ordering the eigenvalues
of a diagonalized matrix at each site.  The anomalous behavior
is intimately connected with this freedom.  We do not find a
completely satisfactory solution,  nevertheless we
are led to an acceptable definition of sources. In Sec.~\ref{sec:5} we
present the results of our simulation.

\section{Definition of the composite adjoint field}\label{sec:2}

The Abelian projection requires a gauge condition that breaks the 
SU(2) gauge invariance but preserves
U(1) gauge invariance.   
This can be accomplished through the following
two-step construction:
\begin{enumerate}
\item   Consider a path ordered product of links, i.e. a Wilson line,
 which begins and ends at 
a particular site  $x$.  This defines an SU(2) group element, $W(x)$, e.g.,
an open plaquette or Polyakov line  at 
the site $x$ (no final trace at $x$).  
This in turn defines a composite field, $\phi(x)$.  
\begin{eqnarray}
W(x) = \cos{\theta(x)} +  i \phi(x) \sin{\theta(x)}
, \;\;\;\phi(x) = \widehat{\phi}(x) \cdot \vec{\sigma} ,\;\;\;
\widehat{\phi}(x) \cdot  \widehat{\phi}(x) = 1, \;\;\;
     \theta \in [0,\pi].
\label{e2.1}
\end{eqnarray}
This composite field 
transforms under the adjoint representation of SU(2),
\begin{eqnarray} 
 \phi(x)  \rightarrow \tilde{\phi}(x) = g(x) \phi(x) g^{\dagger}(x).
 \label{2.101}
 \end{eqnarray}

We require two further generalizations to cover the adjoint 
field
definitions used in this paper.
(i) In the SU(2) fundamental representation, a sum of group elements is,
up to a normalization, also a group element.  Hence we can also construct
an adjoint field from a sum of the above defined 
Wilson lines.  (ii) Further, the
gauge invariant path ordered products of links can include the 
adjoint field itself, $\phi(x')$,  at sites  along the path. This 
leads to a self consistent definition of an adjoint field as, e.g., 
 the case of the maximal Abelian gauge.

\item  After constructing $\phi(x)$, we then 
perform a gauge transformation at each $x$ that fixes the adjoint field 
$\widehat{\phi}(x) $  in the  $3$ direction 
\begin{eqnarray}
 \widehat{\phi}(x)   \rightarrow (0,0, \pm 1), \;\;\;\; \forall x.
\label{e2.101}
\end{eqnarray}
This gauge fixing construction is invariant under subsequent U(1)
gauge transformations which rotate in the $(1,2)$ plane, leaving the
$3$ direction invariant.  
\end{enumerate}

The ambiguity of rotating 
$ \widehat{\phi}(x)$  into the $+ $ or $-$ $3$-direction 
is equivalent to the choice  of ordering the eigenvalues of  
$\widehat{\phi}(x) \cdot \vec{\sigma} $.  
 There are $ 2^N $ variants on 
gauge fixing corresponding to choosing $+1$ or $-1$ in the 3-direction
at each of the N sites.  We address this issue in Sec. IV.

\subsection{Polyakov Gauge}

The Polyakov gauge is defined by the above construction in which $W(x)$
is taken to be an open Polyakov line, $P(x)$
 beginning and ending at $x$.
\begin{eqnarray}
 P(x) = \cos{\theta(x)} +  i \phi_{P}(x) \sin{\theta(x)}
, \;\;\;\phi_P(x) = \widehat{\phi}_{P}(x) \cdot \vec{\sigma} ,\;\;\;
\widehat{\phi}_{P}(x) \cdot  \widehat{\phi}_{P}(x) = 1, \;\;\;
     \theta \in [0,\pi].
\label{e2.1a}
\end{eqnarray}
In Section IV we argue there that we should choose  
$\widehat{\phi}_P(x) \rightarrow (0,0,+1)$  at all sites, or
equivalently $(0,0,-1)$ at all sites.

\subsection{Maximal Abelian gauge}

\subsubsection{Formulation in terms of an adjoint composite field}

The maximal Abelian gauge\cite{klsw}
 is conventionally defined as a configuration, $\{U_{\mu}(x)\}$, 
 which maximizes  ${\cal R}[U_{\mu}(x)]$, 
\begin{eqnarray}
{\cal R}[U_{\mu}(x)] = \sum_{x, \mu} 
\frac{1}{2}Tr\biggl[\sigma_3 U_{\mu}(x)\sigma_3 
U_{\mu}^{\dagger}(x)\biggr],
\label{e2.16}
\end{eqnarray}
over the set of gauge equivalent configurations 
$\{U_{\mu}(x)\}$.  The continuum limit of this condition is
\begin{eqnarray} 
\left(\partial_{\mu} \pm i g A^3_{\mu}\right) 
\left[ A^1_{\mu} \pm  i A^2_{\mu}\right] = 0.
\end{eqnarray}

We wish to cast this into the
language of the composite adjoint field.  To do this let us 
consider a gauge fixing sweep in which we propose a gauge 
transformation on the links $U_{\mu}(x) \rightarrow 
U_{\mu}'(x) = g(x) U_{\mu}(x) g^{\dagger}(x + \mu)$ to test
for an increase in ${\cal R}$.  Then  ${\cal R}$ becomes
\begin{eqnarray}
{\cal R}[U_{\mu}'(x)] &=& \sum_{x, \mu} 
\frac{1}{2}Tr\biggl[\sigma_3 U_{\mu}'(x)\sigma_3 
U_{\mu}'^{\dagger}(x)\biggr],
\nonumber \\
 &=& \sum_{x, \mu} 
\frac{1}{2}Tr\biggl[\sigma_3 \biggl(g(x) U_{\mu}(x) g^{\dagger}(x + \mu)
\biggr)\sigma_3 
\biggl(g(x + \mu) U_{\mu}^{\dagger}(x) g^{\dagger}(x) \biggl)\biggr],
\nonumber \\
 &=& \sum_{x, \mu} 
\frac{1}{2}Tr\biggl[\biggl(g^{\dagger}(x)\sigma_3 g(x)\biggr)
 U_{\mu}(x) \biggl(g^{\dagger}(x + \mu)
\sigma_3 
g(x + \mu)\biggr) U_{\mu}^{\dagger}(x)  \biggr].
\label{e2.16a}
\end{eqnarray}
Hence a proposed gauge transformation of $U$ is equivalent to proposing
a change in the component of $\vec{\sigma}$ from $\sigma_3$ at each site to  
$\phi(x) = \widehat{\phi}(x) \cdot \vec{\sigma}$, 
while leaving $U$ unchanged.

Let us reformulate this procedure slightly
by defining a gauge invariant quantity
${\cal S}$ similar to  ${\cal R}$  but in which  $\sigma_3$
is promoted to a variable 
$\phi(x) = \widehat{\phi}(x) \cdot \vec{\sigma}$ transforming
as Eqn.(\ref{2.101}).
\begin{eqnarray}
  {\cal S}[\phi(x)] = \sum_{x, \mu} 
\frac{1}{2}Tr\biggl[\phi(x)U_{\mu}(x)\phi(x+\mu)
U_{\mu}^{\dagger}(x)\biggr].
\label{e2.15}
\end{eqnarray}
We search for the maximum by proposing  $\{\phi'(x)\}$,
while holding $\{U_{\mu}(x)\}$ constant,  to test for
an increase in ${\cal S}$ .  

Having found the maximum, if  we then  perform
a guage transformation on both $\phi$ and $U$, which rotates 
$ \widehat{\phi}(x)   \rightarrow (0,0, + 1)$ , or $(0,0,-1)$ 
for all $x$, then  
${\cal S} \rightarrow {\cal R}$. 
But ${\cal S}$  is gauge invariant by construction and hence 
in this last step,  the value of ${\cal S}$ does not change,
and the transformed links are the solution to the maximization problem,
 Eqn.(\ref{e2.16}).

This construction divides the  maximal Abelian 
gauge fixing procedure into two logically independent
steps described above, unifying maximal Abelian gauge fixing
with other gauge fixing schemes.

\subsubsection{Stationary condition}

In order to find the set $\{\phi(x)\}$ corresponding to the maximum
of ${\cal S}$,
consider variations of $\cal S$ about the stationary point under an
infinitesimal transformation of $\phi(x)$, while leaving 
$\{U_{\mu}(x)\}$ constant:
\begin{eqnarray}
 \phi(x) \rightarrow \tilde{\phi}(x) &=& 
 \exp \{-i \vec{\eta}(x) \cdot \vec{\sigma}\} \;\phi(x) \;
 \exp \{+i \vec{\eta}(x) \cdot \vec{\sigma}\}
 \nonumber \\ 
 &=&
\biggl\{1 - i \vec{\eta}(x) 
\cdot \vec{\sigma} -\frac{1}{2}\eta^2(x)\biggr\}
	\;  \phi(x)  \; \biggl\{1 + i \vec{\eta}(x) \cdot 
	  \vec{\sigma} -\frac{1}{2}\eta^2(x)\biggr\}.
\label{e2.17}
\end{eqnarray}
The condition that ${\cal S}$ is stationary under these variations is
\begin{eqnarray}
 Tr\biggl([\vec{\eta}(x) \cdot \vec{\sigma}, \phi(x)] \Phi(x)\biggr) = 0,
 \;\;\;\; \forall x,
\label{e2.18}
\end{eqnarray}
where $\Phi(x)$ is defined
\begin{eqnarray}
\Phi(x) 
%\equiv \sum_{\mu}  \phi(x + \mu)|_{p.t.} + \phi(x - \mu)|_{p.t.}
\equiv \sum_1^8 \phi(x)|_{p.t.}.
\label{e2.19}
\end{eqnarray}
The notation $|_{p.t.}$ means that the quantity is parallel transported
to the point $x$ as indicated in Fig.~\ref{f1.0}.
The sum is over the nearest neighbor sites to the point
$x$.    Writing
$\Phi(x) = \vec{\Phi}(x) \cdot \vec{\sigma} $, the stationary condition,
Eqn.~(\ref{e2.18}), gives: 
\begin{eqnarray}
 \vec{\eta}(x) \cdot\widehat{\phi}(x) \times 
\vec{\Phi}(x) = 0, \;\;\;\;  \forall x. 
\label{e2.195}
\end{eqnarray}
Since $\vec{\eta}(x)$ is arbitrary, 
 $\widehat{\phi}(x)$ must be either parallel or 
 antiparallel to $\vec{\Phi}(x)$
 at each site. Therefore
\begin{eqnarray}
 \epsilon(x) \widehat{\phi}(x) = 
 \frac{\vec{\Phi}(x)}{\bigg| \vec{\Phi}(x) \bigg|}
 = \widehat{\Phi}(x),
      \;\;\;\; \forall x,
\label{e2.2}
\end{eqnarray}
where $\epsilon(x) = \pm 1$. 

Let us now examine the second derivatives of $\cal S$ about the
stationary point.  
Define
\begin{eqnarray}
   \Delta{\cal S} =  {\cal S}[\tilde{\phi}(x)] - {\cal S}[\phi(x)].
\label{e5.25}
\end{eqnarray}
Then, using the stationary condition, Eqn.~(\ref{e2.2}), and keeping 
second order terms
\begin{eqnarray}
  \Delta{\cal S}  \approx 
\frac{1}{2}\sum_{x} \biggl[- \eta^2(x)  
\frac{1}{2} Tr\biggl( \phi(x) \Phi(x)\biggr) 
+  \frac{1}{2} Tr\biggl(\vec{\eta}(x) \cdot \vec{\sigma} \;\;
 \phi(x) \;\;\vec{\eta}(x) \cdot \vec{\sigma} \;\;
\Phi(x)\biggr)\biggr].
\label{e5.3}
\end{eqnarray}
This can be cast into the form
\begin{eqnarray}
  \Delta{\cal S} \approx    -\sum_{x}\epsilon(x) 
|\vec{\Phi}(x)|(\vec{\eta}(x) \times \widehat{\phi}(x))^2.
\label{e5.4}
\end{eqnarray}
This shows that the second derivative with respect to the
variables  $\{\eta(x)\}$ is 
 $\epsilon(x) \times$ a negative number. 
If we choose $\epsilon(x) = +1$ for all $x$, then ${\cal S}$
is a local maximum under these infinitesimal variations.
 This defines the composite
adjoint field  $\phi_{M A}(x)$, i.e. step (1) in the process.  

In step (2) we perform a gauge transformation which takes 
$ \widehat{\phi} $ into the 3-direction.  
If we choose a
gauge transformation
that brings  $\phi \rightarrow + \sigma_3$ at all sites, 
then ${\cal S}$ is cast
into the form ${\cal R}$.  This is the transformation to the
the maximal Abelian gauge, and justifies our notation: 
$\phi_{M A}(x)$.  Equivalently, we can take 
$\phi \rightarrow - \sigma_3$ at all sites and reach the
same conclusion.

Consider a solution corresponding to $\epsilon(x) = +1, \forall x $.
If we  then set  $\epsilon(x') \rightarrow  -1$ at one site, $x'$,
 and search anew for a solution, we will  then obtain
one  in which $\widehat{\phi}$ is given by the normalized 
average of $\widehat{\phi}$ at the neighboring sites, parallel 
transported to $x$, except that $\widehat{\phi}(x')$ is  antiparallel
to the average of its neighbors.  
Equation(\ref{e5.4}), shows that the new solution is  
no longer a local maximum of ${\cal S}$ and hence it
is not  associated with the maximal Abelian gauge. We will return
to this issue in a discussion of the minimal Abelian gauge in 
Sec. IV.

In order to find a  stationary 
solution one can start with
an arbitrary configuration $\{\phi(x)\}$ and then iterate. 
 To update the site
$x$, one calculates $\widehat{\Phi}(x)$, Eqn.(\ref{e2.19}), 
based on the current
neighboring $\phi$'s.  Then set 
\begin{eqnarray}
 \biggl[\widehat{\phi}(x) +
\zeta ( \widehat{\Phi}(x) - 
 \widehat{\phi}(x) ) \biggr]_{normalized} 
  \;\; \rightarrow  \;\; \widehat{\phi}(x),
\end{eqnarray}
where $\zeta$ is the over-relaxation parameter.  For $\zeta = 1$ we
go directly to the stationary point for that site based on the
current values of the neighboring $\widehat{\phi}$'s.  However by employing
over-relaxation with $\zeta \approx  1.7 $, the overall search rate for
a self consistent solution is greatly enhanced.  We tune
this parameter to optimize performance.

This procedure requires an iterative search for a 
solution but it does not require iterative gauge transformations.
After finding a solution, a single gauge transformation brings the
gauge field into the maximal Abelian gauge.  
 This method
of obtaining the maximal Abelian gauge  involves fewer multiplications
 than directly maximizing 
${\cal R}$ through gauge transformations and hence is a little faster.

 The maximal Abelian gauge condition has multiple
solutions, known as Gribov copies\cite{gribov} 
for a given field configuration. 
 The particular solution depends on
the initial configuration and on the update algorithm. However
Bali et. al.\cite{bbmps} have studied these 
multiple solutions and have offered
an algorithm that limits this ambiguity.  This Gribov ambiguity
does not seem to have a large effect on the measured values of
Abelian projected observables.  In our discussion of
the maximal Abelian gauge, we should understand that the above 
procedure is very unlikely to find the global maximum since there is no
direct procedure to achieve that.  Hence it must
 be understood that in general  we
are only at a local maximum, not the global maximum, of ${\cal R}$.

\section{Gauge invariant definition of Abelian field strength}
\label{sec:3}

The lattice Georgi-Glashow model\cite{gg} has 
an elementary adjoint Higgs
field living on the sites coupled to the gauge field living on the
links.   This allows 
 one to define a gauge invariant field strength.
The lattice definition of this is given in Fig.~\ref{f2.0}. 
This is the construction used to identify the ordinary magnetic 
field due to monopoles in non-Abelian gauge theory\cite{thooft2}.  
The classical continuum limit of these operators gives:
\begin{eqnarray}
 P_{\mu \nu} \approx&& - iga^2 Tr\biggl[ \phi \biggl(\partial_{\mu} A_{\nu}
                              -  \partial_{\nu} A_{\mu}
                          - ig[A_{\mu}, A_{\nu}] \biggr)\biggr]
\nonumber\\
 &&+
\frac{1}{2} a^2 Tr\biggl[\phi 
\biggl(\partial_{\mu} \phi - ig[A_{\mu},\phi]\biggr)
 \biggl(\partial_{\nu} \phi - ig[A_{\nu},\phi]\biggr)\biggr],
\label{e3.1}
\end{eqnarray}
where $A_{\mu}$ is defined $U_{\mu} = \exp{(-iga A_{\mu})}$, $a$ is the 
lattice spacing and $g$ is the gauge coupling.  This manifestly 
gauge invariant form can be cast into the alternative form: 
\begin{eqnarray}
 P_{\mu \nu} \approx -iga^2  F_{\mu \nu} =
 -iga^2 \biggl[ \partial_{\mu} (\widehat{\phi}\cdot \vec{A}_{\nu})
                        - \partial_{\nu} (\widehat{\phi}\cdot \vec{A}_{\mu})
    - \frac{1}{g} \widehat{\phi} \cdot (\partial_{\mu} \widehat{\phi}) \times
                                   (\partial_{\nu} \widehat{\phi}) \biggr],
\label{e3.2}
\end{eqnarray}
where we have used  $\phi = \widehat{\phi} \cdot \vec{\sigma}$, 
($ \widehat{\phi} \cdot \widehat{\phi} = 1 $), and
$A_{\mu} = \vec{A}_{\mu} \cdot \vec{\sigma}/2$.  

We can implement a gauge transformation that brings 
$ \widehat{\phi} \rightarrow (0, 0, 1)$ 
for all $x$.  Then the last term of Eqn.~(\ref{e3.2}) vanishes.  
The first term gives a conventional Abelian field strength, 
\begin{eqnarray}
 F_{\mu \nu} = \partial_{\mu} W_{\nu} -  \partial_{\nu} W_{\mu} ; \;\;\;\;
W_{\mu} \equiv  (\widehat{\phi}\cdot \vec{A}_{\mu}) = (\vec{A}_{\mu})_3.
\label{e3.3}
\end{eqnarray}

Whether $\widehat{\phi}$ denotes an elementary Higgs field, or a
composite adjoint field as we discussed in 
Sec. II, this definition of
field strength, 
Eqn.~(\ref{e3.1}), is a manifestly gauge invariant quantity.
However this is perhaps misleading in the composite adjoint field
case because the definition of the field is identified with
a specific gauge choice.

Del Debbio et al. \cite{ddpp} employed $\widehat{\phi}_P$ in defining the
Abelian field strength.  This was needed in order to construct the 
monopole field operator.   The vacuum expectation value of this field
gave a very clear signal for dual superconductivity.   
Although their construction
is manifestly gauge invariant,  it is identified with the Polyakov gauge
in the sense of this section.   It is this connection that prompted us
to study vortex formation in the Polyakov gauge to probe the connection
between vortices and  a non-zero vacuum expectation value.

\section{Adjoint field and the Abelian projection}\label{sec:4}

The Abelian projection consists of (i) choosing a gauge  that preserves
a U(1) gauge invariance,  and
(ii) identifying Abelian links associated
with that residual invariance\cite{ksww}.  
In Sec. II we  discussed the first step of choosing the gauge 
except for the ambiguity of transforming  $\widehat{\phi}$ into 
the $ + 3$ or $ - 3$-direction at each site. 
 We will argue here that the $+ 3$-direction
is the preferred choice for all sites or $ - 3$ for 
all sites.  However this leads to an unsettling result for
the Polyakov gauge as we point out in part B below.  
Hence we are led to examine the issue more
broadly to check out alternatives in the following parts of this
section.  There are $2^N$ choices, where $N$ is the number of sites.

The choice of rotating $\widehat{\phi}$ into the + 3-direction 
at every site is the prescription one would anticipate  in order to get a 
smooth classical limit in which the changes in the gauge
field are small between one site and its neighbors.   But we
feel alternatives should be presented for the record.   The
reader can skip to Part H, the conclusion for this section, 
without losing continuity.
We have summarized the key results of this 
section in Table~\ref{t1}.

\subsection{Abelian links}

For the purpose of defining Abelian links, consider the 
parametrization\cite{cpv2}:
\begin{eqnarray}
 U_\mu (x) &=& \left (
\begin{array}{cc}
\cos (\phi_\mu (x)) e^{{\textstyle i\theta_\mu (x)}} &
\sin (\phi_\mu (x)) e^{{\textstyle i\chi_\mu (x)}} \\& \\
-\sin (\phi_\mu (x)) e^{{\textstyle -i\chi_\mu (x)}} &
\cos (\phi_\mu (x)) e^{{\textstyle -i\theta_\mu (x)}} \\
\end{array}
\right ), 
\label{e4.1} \\
\phi_\mu &\in& \left (0,\frac{\pi}{2}\right ), \;\;\;\;
\theta_\mu \;, \chi_\mu \in (-\pi,\pi). 
\nonumber
\end{eqnarray}
The Abelian link is defined through the factorization of
Eqn.~(\ref{e4.1})
\begin{eqnarray}
 \left (
 \begin{array}{cc}
\cos (\phi_\mu (x))  &
\sin (\phi_\mu (x)) e^{{\textstyle i\gamma_\mu (x)}} \\& \\
-\sin (\phi_\mu (x)) e^{{\textstyle -i\gamma_\mu (x)}} &
\cos (\phi_\mu (x))  \\
\end{array}
\right )
 \left (
\begin{array}{cc}
e^{{\textstyle i\theta_\mu (x)}} &
0 \\& \\
0 &
e^{{\textstyle -i\theta_\mu (x)}} \\
\end{array}
\right ),
\label{e4.2} 
\end{eqnarray}
where $\gamma_\mu (x) = \chi_\mu (x) + \theta_\mu (x)$. The Abelian link
is given by $U_{\mu}(x) = e^{{\textstyle i\theta_\mu (x)}}$. 
The complex coset field is given by
$\sin (\phi_\mu (x)) e^{{\textstyle i\gamma_\mu (x)}}$.
Under a U(1) gauge transformation, $g(x) = e^{i\alpha(x)\sigma_3}$
\begin{eqnarray}
 \theta_{\mu}(x) &\rightarrow& \theta_{\mu}(x) + 
\alpha(x) - \alpha(x + \mu), \\ \nonumber
\gamma_{\mu}(x)  &\rightarrow&   \gamma_{\mu}(x)  + 2 \alpha(x).
\label{e4.3}
\end{eqnarray}
The coset vector fields live on the sites and are doubly charged. 

The ambiguity of rotating $\widehat{\phi}$ to 
the $\pm 3$ direction, $ \phi = \pm \sigma_3$,
is the same as the ambiguity of ordering the
eigenvalues of this $2 \times 2$ diagonal matrix.
The order of the eigenvalues for a particular link can be
interchanged via an SU(2) gauge transformation $g = i\sigma_2$
at the two neighboring sites corresponding to the link in question.

\subsection{The Polyakov gauge with the $+$ prescription}

Translation invariance suggests that whatever rule we adopt for
ordering the eigenvalues, we should use the same rule at every site.  
Let us choose $\phi = + \sigma_3$, denoting this as the $+$ prescription.

In the Polyakov gauge if we
choose $\phi_P = + \sigma_3$ at every site, then there is a 
difficulty.  The SU(2) Polyakov line, Eqn.~(\ref{e2.1}), can be written
\begin{eqnarray}
\left (
 \begin{array}{cc}
e^{i \theta_P(x)}  &   0
 \\& \\
0 &
e^{-i \theta_P(x)}  \\
\end{array}
\right ), \;\;\;\;  \theta_P(x) \in [0,\pi].
\label{4.31}
\end{eqnarray}
The $11$ element of this is the U(1) Polyakov line
$P(x) = e^{i\theta_{P}(x)}$.  Since $\theta_P$ is 
distributed in the domain $[0,\pi]$ the imaginary part of $P(x)$
can not be negative.  As a consequence, the average of $P(x)$ 
will have a non-zero imaginary
part due to this kinematical effect. 
This is anomalous. Polyakov lines represent static sources.  
If we are to get the `expected' 
distribution in which $\theta$ is distributed symmetrically in 
the domain $(-\pi,\pi]$ we would need a rule that allows the
choice $\phi_P = - \sigma_3$ with a statistical weight equal to
the $\phi_P = + \sigma_3$ case.

If we choose $\phi \rightarrow - \sigma_3$ at all sites, we reach
the same conclusion except that $Im(P(x))$ can not be positive.

\subsection{U(1) gauge theory as a guide}

We can examine the U(1) model for guidance.
We expect Abelian projected SU(2) to resemble the U(1) theory. 
Consider the limit in which the SU(2) theory goes over to
the U(1) theory. 
In SU(2) the links lie in the group manifold $S_3$.  In U(1) they
lie in a submanifold $S_1$.  Consider adding a term to
the action that is invariant under U(1) but
which breaks SU(2).   
Further choose it to bias the links into a peaked
distribution centered on $S_1$. 
Then as the width of the peak approaches
zero, the SU(2) theory goes over to the U(1) theory.  
In U(1),  the Polyakov line   
$P(x) = e^{i\theta_P(x)}$ takes values for
$\theta_P$ distributed in the domain $(-\pi,\pi]$.   Positive
and negative values of $\theta_P$ are equally probable. 

\subsection{The $\pm$ prescription $\Longleftrightarrow$ random 
prescription}

To try to resolve the discrepancy between the U(1) theory
and Abelian projected SU(2) in the $+$ prescription, 
let us consider a prescription in 
which we choose $+$ or $-$ depending on the
value of $\widehat{\phi}(x)$.  For example, we can correlate the
signs $+$ or $-$ with the sign of $[\widehat{\phi}(x)]_3$.  In other words, 
if $\widehat{\phi}(x)$ is in the upper hemisphere then choose $+$.  If it
is in the
lower hemisphere then choose $-$.  We denote this the $\pm$ prescription.
The phase of the Abelian 
Polyakov line will now be distributed in the expected way, $(-\pi,\pi]$.
This is perhaps the prescription most closely associated with
 the limiting procedure in the
previous section.

 This prescription is equivalent
to choosing $+$ or $-$ randomly.   The reason is that a 
gauge transformation can give an arbitrary orientation to 
$\widehat{\phi}$.   If after each update, we follow it with a
random gauge transformation, (an update followed by a gauge 
transformation is indistinguishable from
an update alone) then the assignments of
 $+$ or $-$ at each site will be random.  
The SU(2) transformations do not respect the
homotopy classes of U(1).

\subsection{The Polyakov gauge with the $\pm$ prescription}\label{sec:4.d}

We now apply the $\pm$ prescription to the Abelian
projection in the Polyakov gauge and point out a flaw in this
choice.

Let $P$ denote the open Polyakov line at site $x$, and 
$P'$ the open Polyakov line at the neighboring site in the
$+ x_4$ direction.  Then 
\begin{eqnarray}
 P U = U P',
\label{e4.41} 
\end{eqnarray}
where $U$ is the link
in the $x_4$ direction connecting the two sites.  

Now go to
the Polyakov gauge.  Suppose we choose $\phi_P = + \sigma_3$ at 
both sites.  Then $ P = P'$ and the solution 
to Eqn.~(\ref{e4.41}) gives
\begin{eqnarray}
 U  = u_4 + i u_3 \sigma_3,\;\;\; u_3^2 + u_4^2 = 1,
\label{e4.42} 
\end{eqnarray}
showing that the  links in the $x_4$ direction are diagonal.  
Contrast this with the case in which we choose $\phi_P = + \sigma_3$
at one site and $\phi_P = - \sigma_3$ at the neighboring site.  Then
$P' =  P^{\dagger}$ and the solution 
to Eqn.~(\ref{e4.41}) gives
\begin{eqnarray}
 U  = i(u_1 \sigma_1 + u_2 \sigma_2) \;\;\; u_1^2 + u_2^2 = 1. 
\label{e4.43}
\end{eqnarray}
This second case does not give an acceptable Abelian projection since
the $x_4$ Abelian link is 
undefined, i.e. the $11$ element of the link matrix
is zero.  

We can still define 
a $\pm$ prescription appropriately
modified for the Polyakov gauge in which the spatial sites have
independent random $\pm$ choices, but that the choice is independent
of the values of $x_4$.  

However this construction has a further problem.  Consider the correlator
shown in Fig.~\ref{f2.5}. {\it The correlation of the 
field strength with a product of
two Abelian Polyakov lines representing a static quark-antiquark pair 
vanishes.} 
This correlation is given by 
\begin{eqnarray}
 \langle e^{i (\pm \theta_P(x_1) \pm \theta_P(x_3))}  
 \times \sin \theta_F(x_2) \rangle.
\label{e4.431}
\end{eqnarray}
The $\pm$ signs, arising from the prescription of a random
gauge transformation, $g = i\sigma_2$, have the effect of 
flipping the sign of the charge of the two sources.  With a
statistical  weight equal for each of the four signs, 
the field averages to zero.

The correlation also vanishes if one only considers random $ i\sigma_2$ gauge 
transformation at the sites contiguous with the plaquette.
Consider the parametrization of a link in Eqn.~(\ref{e4.1}).   We adopt
this parametrization 
for a link in which the $+$ prescription is applied at both
ends.  The SU(2) link and the corresponding Abelian link are
\begin{eqnarray}
 U^{++}_\mu  = \left (
\begin{array}{cc}
\cos \phi_\mu  e^{{\textstyle i\theta_\mu }} &
\sin \phi_\mu  e^{{\textstyle i\chi_\mu }} \\& \\
-\sin \phi_\mu  e^{{\textstyle -i\chi_\mu }} &
\cos \phi_\mu  e^{{\textstyle -i\theta_\mu }} \\
\end{array}
\right ),
\hspace{0.7cm} U^{Abelian}_\mu = e^{{\textstyle i\theta_\mu }}.
\label{e4.432}
\end{eqnarray}
Different prescriptions will change the SU(2) links and accordingly 
the choice of the Abelian angles. Let $i\sigma_2$ be the gauge 
transformation changing 
the orientation of the adjoint field. (This gauge transformation is  
defined up to a $U(1)$ transformation.) The link will transform according 
to one of the following cases:
\begin{eqnarray}
 U^{+-}_\mu =  U^{++}_\mu (-i\sigma_2) =\left (
\begin{array}{cc}
\sin \phi_\mu  e^{{\textstyle i\chi_\mu }} &
-\cos \phi_\mu  e^{{\textstyle i\theta_\mu }} \\& \\
\cos \phi_\mu  e^{{\textstyle -i\theta_\mu }} &
\sin \phi_\mu  e^{{\textstyle -i\chi_\mu }} \\
\end{array}
\right ),
\hspace{0.7cm} U^{Abelian}_\mu = e^{{\textstyle i\chi_\mu }},
\label{e4.433}     \\ \nonumber \\
 U^{-+}_\mu  = i\sigma_2 U^{++}_\mu =\left (
\begin{array}{cc}
-\sin \phi_\mu  e^{{\textstyle -i\chi_\mu }} &
\cos \phi_\mu  e^{{\textstyle -i\theta_\mu }} \\& \\
-\cos \phi_\mu  e^{{\textstyle i\theta_\mu }} &
-\sin \phi_\mu  e^{{\textstyle i\chi_\mu }} \\
\end{array}
\right ),
\hspace{0.7cm} U^{Abelian}_\mu = e^{{\textstyle i(\pi -\chi_\mu )}}, 
\label{e4.434}     \\  \nonumber   \\
 U^{--}_\mu  =i\sigma_2 U^{++}_\mu (-i\sigma_2) = \left (
\begin{array}{cc}
\cos \phi_\mu  e^{{\textstyle -i\theta_\mu }} &
\sin \phi_\mu  e^{{\textstyle -i\chi_\mu }} \\& \\
-\sin \phi_\mu  e^{{\textstyle i\chi_\mu }} &
\cos \phi_\mu  e^{{\textstyle i\theta_\mu }} \\
\end{array}
\right ),
\hspace{0.7cm} U^{Abelian}_\mu = e^{{\textstyle -i\theta_\mu }}
\label{e4.435}    
\end{eqnarray}
where the $\pm\pm$ notation refers to the prescription at each
end of the link.

After a random gauge transformation the $sin$ of the Abelian plaquette
 angle will assume one of the 
forms shown in Fig.~\ref{f2.6}.  
We are working in the Polyakov gauge hence there are only
four possibilities since the adjoint field must be constant in $x_4$.  
All the above configurations have the same statistical weight and for each
configuration there is always another configuration differing only by 
the sign.  Hence the field strength in the presence of 
two sources vanishes in the $\pm$ prescription according to this argument
also.

This argument generalizes to any Wilson loop since the 
parametrization could apply to a Wilson line instead of a single
link.  Further we can replace the Polyakov line source by another
Wilson loop and then we obtain
\begin{eqnarray}
\langle\sin \theta_{W_1} \sin \theta_{W_2}  \rangle = 0,
\label{e4.436}    
\end{eqnarray}
for the $\pm$ prescription if the two Wilson loops do not share
any spatial sites.

\subsection{Gauge invariant field strength and the $+$ prescription}

To get further support for the $+$ prescription
consider the Abelian projected theory in the Polyakov gauge but 
instead of calculating the field strength from the
Abelian plaquette let us instead represent the field strength by
the manifestly gauge invariant form, Fig.~\ref{f2.0}.  

In the 
Polyakov gauge, $\widehat{\phi}(x)$ is independent of $x_4$ and as
a consequence, the second term in Fig.~\ref{f2.0} vanishes.
Consider a static source
represented by an Abelian Polyakov line at point $x$, and the field 
strength at point $y$ as shown in Fig.~\ref{f3.0}(a). 

Let us define variables for the
(open) plaquette in Fig.~\ref{f3.0}(a) by
\begin{eqnarray}
 \cos \theta_F(y) +
i \vec{\sigma} \cdot \widehat{\phi}_F(y) \sin \theta_F(y).
\label{e4.61} 
\end{eqnarray}
Then the correlator is given by
\begin{eqnarray}
\frac{\langle\sin \theta_P(x)
\widehat{\phi}_P(y) \cdot \widehat{\phi}_F(y)
\sin \theta_F(y)\rangle}{\langle\cos \theta_P\rangle}  = 
\frac{\langle\sin \theta_P(x)
(\widehat{\phi}_F(y))_3
\sin \theta_F(y)\rangle}{\langle\cos \theta_P\rangle},
\label{e4.62} 
\end{eqnarray}
where the last equality is due to the choice of the Polyakov gauge.
Measuring the field strength with the Abelian plaquette, 
Fig.~\ref{f3.0}(b), gives
\begin{eqnarray}
 \frac{\langle\sin \theta_P(x)
\sin \theta^{(Abelian)}_F(y)\rangle}{\langle\cos \theta_P\rangle}.
\label{e4.63} 
\end{eqnarray}
The field strength in Eqn.~(\ref{e4.62}) is manifestly
gauge invariant, and therefore the prescription at the site $y$
has no bearing on the definition of the correlator. 
If we use the Abelian plaquette, Eqn.~(\ref{e4.63}), instead,
then the value depends on the prescription.  However the $+$
prescription has the same classical limit for small lattice spacing
$a$ as the classical limit of the gauge 
invariant form, Eqn.(\ref{e3.3}).  This follows from the discussion
in Sec.~\ref{sec:4.d}.   
In Fig.~\ref{f2.6} in the cases in which the four points
of the plaquette have the same prescription, the Abelian plaquette 
has the standard classical limit.   In the other cases, the plaquette
involves angles arising from the off-diagonal elements of the 
SU(2) link.

For completeness we compare this to the connected definition
of field strength at point $y$ as shown in Fig.~\ref{f3.0}(c). 
In our notation the correlator is given by
\begin{eqnarray}
 \frac{\langle(\sin \theta_P(x)
\widehat{\phi}_P(x))|_{p.t.(x \rightarrow y)} \cdot \widehat{\phi}_F(y)
\sin \theta_F(y)\rangle}{\langle\cos \theta_P(x)\rangle},   
\label{e4.64} 
\end{eqnarray}
where $|_{p.t.(x \rightarrow y)}$ means that the  
variables are parallel transported from $x$ to $y$ in the same sense
as applied to Eqn.(\ref{e2.19}).

\subsection{Minimal Abelian projection}
\label{sec:4.f}

We consider only lattices which allow an even-odd checkerboard 
assignment to the sites.  Consider two solutions of
Eqn.~(\ref{e2.2}), corresponding to different values of
$\epsilon(x)$.
\begin{eqnarray}
\widehat{\phi}_{MA}(x)  &:&  
  \;\;\;\; \epsilon(x) = +1 \;\;\forall\;\; x, \nonumber \\
\widehat{\phi}_{mA}(x) &:& 
   \;\;\;\; \epsilon(x) = -1 \;\;\forall \;\; x. 
\label{e5.411}
\end{eqnarray}
For the case in which the LHS of Eqn.~(\ref{e2.2}) 
involves only odd  (even) 
sites then the  RHS involves only even (odd) sites, and vice versa.    
Hence 
\begin{eqnarray}
\widehat{\phi}_{mA}(x)  &=&   + \widehat{\phi}_{MA}(x), 
\;\; x \; \textstyle{odd (even)}, \nonumber \\
 &=&   - \widehat{\phi}_{MA}(x), 
\;\; x \; \textstyle{even (odd)}.
\end{eqnarray}
In other words, the $\widehat{\phi}_{mA}(x)$ solution is identical 
to the $\widehat{\phi}_{MA}(x)$ solution
 except that it is flipped at the 
even (odd) sites. This is in contrast to the example in 
Sec. IIB where a change in $\epsilon(x)$ led to a new solution which
has no simple relation to the old solution.
 We further note that each term in
 ${\cal S}[\phi(x)]$,
Eqn.(\ref{e2.15}), involves $\widehat{\phi}(x)$ at an even
site and $\widehat{\phi}(x)$ at an odd site.  Therefore the
effect of the flipping is to change the sign of ${\cal S}$
\begin{eqnarray}
 {\cal S}[\widehat{\phi}_{mA}] = 
       -{\cal S}[\widehat{\phi}_{MA}]
 \label{e4.641} 
\end{eqnarray}
Since  $\widehat{\phi}_{MA}$ maximizes  ${\cal S}$, it follows that
$\widehat{\phi}_{mA}$ minimizes  ${\cal S}$.  This is corroborated
by the stability analysis, Eqn.(\ref{e5.4}).

The adjoint field $\widehat{\phi}_{mA}$ defines the minimal
Abelian projection introduced by Chernodub, Polikarpov and 
Veselov\cite{cpv2}.  They showed that confinement is due
to objects other than monopoles, which they denoted 
as minopoles.

\subsection{Conclusion about the prescriptions}\label{sec:4.g}

The arguments in this section lead us to adopt the $+$ prescription 
for Abelian projected SU(2) 
in the Polyakov gauge.  This prescription is not completely 
satisfactory.  It does not
 go over smoothly to a U(1) theory as one might expect.  This is 
indicated
by the behavior of the phase angle 
of the Abelian projected Polyakov line.  It
is bounded 'artificially' 
in the interval $[0,\pi]$ giving  the expectation
value of the Polyakov line an anomalous imaginary part.  
The adjoint field defined for the maximal Abelian gauge is
a construct that does relate directly to an observable in this
sense and does not seem to cause a similar difficulty.

Further we must rethink how to
represent static sources.  The distribution of the Abelian Polyakov 
phase angle in the domain $[0,\pi]$  is weighted by the
Haar measure, $\sin^2 \theta$ which is peaked at $\pi/2$.
This means, e.g. 
that the average of two Polyakov lines
\begin{eqnarray}
 P_1 P^*_2 = e^{i(\theta_1 - \theta_2)},
\label{e4.44} 
\end{eqnarray}
will have a non-vanishing real part arising from the same kinematic
effect.  Our solution to this is to define the quark-antiquark
source as the connected part of the product of the two Polyakov
loops, 
\begin{eqnarray}
 \langle P_1 P^*_2\rangle - 
 \langle P_1\rangle\langle P^*_2\rangle.
\label{e4.45} 
\end{eqnarray}

\section{Dual vortices and Abelian projections}\label{sec:5}

\subsection{Measurements}
In this section we  study the electric flux and the magnetic monopole 
current distributions around static sources in the Polyakov gauge with the 
$+$ prescription and the definition of  Abelian static sources as in 
Sec.~\ref{sec:4.g}. Moreover we make a comparison between this projection and 
the maximal Abelian projection in which the main properties of dual vortices 
have already been studied.

Our simulations were performed on a $12^3\times 4$ lattice with the standard 
form of the Wilson action. The calculation follows the method given in 
\cite{sbh}.
We used two Abelian Polyakov lines correlated 
along the 3-direction as static sources in the finite temperature theory. 
The longitudinal  electric field 
between two charges separated by a distance $d$ in the 
3-direction is given by:
\begin{equation}
\langle E_z (x)\rangle =\frac{\langle P(0)P^\dagger (d) \sin \theta_{34}(x)
\rangle}{a^2 e \langle P(0) P^\dagger (d)\rangle}=
\frac{\langle \sin \theta_{PP^\dagger}\sin \theta_{34}(x)\rangle}
{a^2 e\langle\cos\theta_{PP^\dagger}\rangle}
\end{equation}
where $\theta_{34}$ is the angle of the Abelian plaquette in the $3-4$ plane, 
$\theta_{PP^\dagger}$ is the angle of the product of the two Abelian 
Polyakov lines and $e$ is the electric charge.

%\noindent
Magnetic monopole currents can be identified by the dual Maxwell 
equations that in the continuum are:
\begin{equation}
J^m_\mu = -\frac{1}{2}\epsilon_{\mu\nu\rho\sigma}\partial_\nu F_{\rho\sigma}
\end{equation}
On the lattice one can choose \cite{zfks}: 
\begin{equation}
a^2 e F_{\mu\nu} = \sin \theta_{\mu\nu} 
\end{equation}
We measured the magnetic current distribution around static sources:
\begin{equation}
\langle({\rm curl}\vec{J}^m (x))_z \rangle = \frac{2\pi \langle \sin 
\theta_{PP^\dagger} ({\rm curl}\vec{J}^m (x))_z\rangle}
{a^4 e \langle\cos\theta_{PP^\dagger}\rangle} 
\end{equation}
where the curl of the magnetic current is constructed as in \cite{sbh}, and 
$2\pi/e$ is the magnetic charge.

\subsection{Numerical results and discussion}
We made our simulations at $\beta =2.25$ and $\beta =2.40$. As we chose the 
temporal extent of the lattice equal to 4, the first value of $\beta$ 
corresponds to the confined phase while the second corresponds to the 
unconfined phase.

The results for the Polyakov gauge are shown in Fig.~\ref{EMpoly} for 
$d=1a,2a,3a$. For $\beta =2.25$ we see a clear signal for both the electric 
field and the curl of the magnetic currents on the axis of the two charges. 
As we move in the transverse direction $r$, we observe a rapid fall off of 
the electric field that always vanishes after a few lattice spacings and  
a behavior of the monopole current consistent with the predictions of the 
Ginzburg-Landau theory. As the distance $d$ of the static charges increases, 
the values of the electric field decreases as expected. 

In the unconfined phase ($\beta =2.40$) we see a remarkably 
different behavior of the curl of the magnetic currents that is much smaller 
than in the other phase and at $d=3a$ is consistent with zero everywhere. 
Moreover the electric field approaches zero less rapidly  although not 
clearly evident from the figure.

A similar scenario has been found for the maximal Abelian gauge\cite{sbh}. In 
Fig.~\ref{EMmag} we show the results in this gauge for the case $d=3a$. 
We used the procedure described in Sec.~\ref{sec:2}  to define the Abelian 
projection.

Although both Abelian projections produce results consistent with the 
Ginzburg-Landau theory, the dual superconductivity parameters are quite 
different for the two projections. We are not able to perform a quantitative 
analysis in order to evaluate the coherence length $\xi_d$ and the London 
penetration length $\lambda_d$. Nevertheless while the data for the maximal 
Abelian gauge support a non-vanishing value for the coherence length, there is
no evidence for $\xi_d$ being different from zero in the Polyakov gauge. 

Moreover a comparison between the two different projections shows  that in 
the Polyakov gauge the peak values of the electric field and of the curl of 
the magnetic currents are more than an order of magnitude smaller than the 
correspondent values in the maximal Abelian gauge.  Suzuki 
et. al.\cite{simoy} reported similarly that the string tension was
suppressed in the Polyakov gauge.

To understand these differences we remember that after Abelian projection 
we are left not only with the Abelian links but also with the  
coset vector fields. The dynamics of these doubly charged fields is 
clearly dependent on the choice of the Abelian projection. 
To show the different role of the coset fields in the two projections, we 
measure the divergence of the electric field giving the spatial distribution 
of the electric charge. The divergence at site $x$ is defined as
\begin{equation}
\langle {\rm div}\vec{E}(x)\rangle = \sum_{i=1}^3\frac{\langle \sin 
\theta_{PP^\dagger}(\sin\theta_{i4}(x)-\sin\theta_{i4}(x-i))\rangle}
{a^3 e \langle\cos\theta_{PP^\dagger}\rangle}
\end{equation}
so that all the six plaquettes have in common the link starting at $x$ 
and extending in the time direction. In Fig.~\ref{Gauss} we show the 
results for two Polyakov lines separated by three lattice spacings. The 
figure clearly shows that the effective charge of the static source 
is much smaller in the  
Polyakov gauge and that the coset fields in the two prescriptions respond in 
opposite ways to the presence of an electric charge with the fields in the 
Polyakov gauge shielding the static charge. Skala et al. have studied 
${\rm div} \vec{E}$ in SU(3) in 
a gauge invariant formulation and conclude there 
is no screening \cite{sfz}.

\section{Conclusions}

Using the  U(1) theory as a guide, we explore the connection between
(i) the non-vanishing of a vacuum expectation value of the monopole field
and (ii) vortex formation.   Item (i) provides the underlying
principle, item (ii) gives direct evidence of a photon mass and confinement.
In the Abelian projection of SU(2) $\rightarrow$ U(1) there is an
additional consideration.  After identifying the U(1) gauge field, 
there remain dynamical charged vector coset fields.   They can screen
the static sources and change the electric flux, affecting the 
string tension.   

From prior work we know that, e.g., the minimal Abelian
projection and the maximal Abelian projection account for the
physics of confinement through very different mechanisms.   
We anticipated that the monopoles in the
Polyakov projection would give a picture very similar to the 
maximal Abelian gauge.  However we found that the coset fields greatly
suppress the static sources.  It could be that we are much farther from 
the continuum limit than we thought.  But that seems unlikely  since
other quantities are close to scaling values.   Our definitions of
field strength,  monopoles, static sources 
etc. have some leeway,  but all should give the
same continuum limit.  Or it could be that the dynamical variables 
arising from the Polyakov Abelian 
projection do not adequately separate the short distance and long
distance physics.   This leaves the maximal Abelian gauge as
the prime candidate  to define an effective theory
of confinement in this scenario.

\acknowledgments

We would like to thank Adriano Di Giacomo and Tsuneo Suzuki for 
discussions.   Work was partially supported by the Department
of Energy under Grant No. DE-FG05-91ER40617.  Giuseppe Di Cecio
was partially supported by a University of Pisa postdoctoral  
fellowship.

\begin{figure}
\caption{Parallel transport of adjoint field. The 
         lines represent single links, $U$. 
         The adjoint field, indicated by the circle, 
         is normalized: $\phi^2 = I, \;\;\;\; 
         \phi = \widehat{\phi} \cdot \vec{\sigma}$.} 
\label{f1.0}
\end{figure}
\begin{figure}
\caption{Definition of gauge invariant Abelian field strength. The 
         lines represent single links, $U$. 
         The adjoint field, indicated by the circle, 
         is normalized: $\phi^2 = I, \;\;\;\; 
         \phi = \widehat{\phi} \cdot \vec{\sigma}$.} 
\label{f2.0}
\end{figure}
\begin{figure}
\caption{Vanishing of the field strength in the presence of
Polyakov lines. } 
\label{f2.5}
\end{figure}
\begin{figure}
\caption{Cancellation of terms for the field
strength in the $\pm$ prescription.} 
\label{f2.6}
\end{figure}
\begin{figure}
\caption{Correlators for measuring field strength in the
neighborhood of a static source.  (a) The form
using the gauge invariant
field strength, (b) using the Abelian plaquette, and 
(c) the connected correlator.} 
\label{f3.0}
\end{figure}
\begin{figure}
\caption{Electric field  and curl of monopole currents as a function 
of the transverse distance $r$ from the axis of two static sources 
for the Polyakov gauge. For separation $d=3a$ the signal in the 
region midway between the charges is drowned out by the noise, 
therefore we only show the result for the region close to one of 
the charges (z=0).}
\label{EMpoly}
\end{figure}
\begin{figure}
\caption{Electric field and curl of magnetic currents as a 
function of $r$ for the maximal Abelian gauge.}
\label{EMmag}
\end{figure}
\begin{figure}
\caption{Divergence of the electric field along the axis connecting 
two static charges separated by three lattice spacings.}
\label{Gauss}
\end{figure}
%

%----------------------------------------------------------------------------
\begin{table}[p]
\begin{tabular}{lcll} \hline
Gauge & P & Definition & Comments 
\\
Condition &  &&
\\ 
\hline \hline
 & ``$+$"  : &&
\\
&$\widehat{\phi}(x) \rightarrow (0,0,+1),\forall x$; &
&
\\ 
 & ``$-$" : &&
\\
&$\widehat{\phi}(x) \rightarrow (0,0,-1),\forall x$; &
&
\\ 
& ``$\pm$": &&
\\
&
$\widehat{\phi}(x) \rightarrow (0,0, \textstyle{sgn}\{{\widehat{\phi}(x)_3}\})$; 
&
&
\\ 
& ``random": &&
\\
&
$\widehat{\phi}(x) \rightarrow (0,0, \textstyle{random \; \pm 1} ).$ 
&
&
\\ \hline \hline
Polyakov & ``$+$" or ``$-$" &  $P(x) = \cos \theta +$&
 Im[Abelian Polyakov line] $\neq0$. 
 \\
&&$ i \widehat{\phi}_P(x) \cdot \vec{\sigma} 
\sin \theta(x)$, & Classical limit of field strength
\\
& & Eqn.(\ref{e2.1a}).
 &  same as classical limit of gauge 
 \\
&&&invariant field strength.
 \\ \hline
 & ``$\pm$"$^1$ & 
 &   Im[Abelian Polyakov line] $ = 0$. 
 \\
& & 
 & Correlations vanish.  
 \\
&&& Time-like Abelian 
\\
&&& links undefined.$^1$ \\ \hline
 & ``random"$^1$ & 
 &  Equivalent to ``$ \pm $''.$^1$ 
 \\ \hline \hline
Maximal & ``$+$" or ``$-$" & $\widehat{\phi}_{MA}(x) 
= +  \widehat{\Phi}_{MA}(x), \; \forall x$,& 
Maximizes ${\cal R},{\cal S},$\\
Abelian &&
 Eqn.(\ref{e2.2}),$\epsilon(x) = +1$.
 & Eqns.(\ref{e2.16}), (\ref{e2.15}).\\  \hline
Minimal & ``$+$" or ``$-$"& $\widehat{\phi}_{mA}(x) 
= - \widehat{\Phi}_{mA}(x),\; \forall x$,& 
Minimizes ${\cal R},{\cal S},$\\
Abelian &&  Eqn.(\ref{e2.2}), $\epsilon(x) = +1$.& 
Eqns.(\ref{e2.16}), (\ref{e2.15}).\\  \hline
\end{tabular}
\caption{Summary of gauge fixing. (1. The problem in defining
time-like Abelian links can be remedied by choosing either 
``$+$" or  ``$-$" at all sites along a particular Polyakov
line.)}
\label{t1}
\end{table} 
%----------------------------------------------------------------------------
%----------------------------------------------------------------------fig 1
\newpage
\begin{figure}[htb]
\setlength{\unitlength}{.3mm}
\begin{picture}(400,160)(30,0)
\newsavebox{\sqccc}
\savebox{\sqccc}{
\put(20,26){\line(1,0){100}}
\put(20,20){\line(1,0){100}}
\put(116,23){\circle*{10}}}
\put(420,80){\usebox{\sqccc}}
\put(410,100){$x$}
\newsavebox{\sqcccc}
\savebox{\sqcccc}{
\put(20,26){\line(1,0){100}}
\put(20,20){\line(1,0){100}}
\put(20,23){\circle*{10}}}
\put(400,10){\usebox{\sqcccc}}
\put(535,30){$x$}
\put(70,30){$ \phi(x)|_{p.t.} = 
U^{\dagger}_{\mu}(x-\mu)\phi(x-\mu)
U_{\mu}(x-\mu)
$}
\put(70,100){$ \phi(x)|_{p.t.} = 
U_{\mu}(x)\phi(x +\mu)U^{\dagger}_{\mu}(x) $}
\end{picture}
\end{figure}
\vspace{6truecm}
\begin{center}
FIG. 1
\end{center}
%---------------------------------------------------------------fig 2
\newpage
\begin{figure}[htb]
\setlength{\unitlength}{.3mm}
\begin{picture}(400,160)(30,0)
\newsavebox{\sqc}
\savebox{\sqc}{
\put(20,20){\line(1,0){100}}
\put(20,120){\line(1,0){100}}
\put(20,20){\line(0,1){100}}
\put(120,20){\line(0,1){100}}
\put(20,20){\circle*{10}}}
\put(90,0){\usebox{\sqc}}
\put(30,70){$ P_{\mu \nu}  \hspace{.2in} = $}
\put(250,70){$+ \hspace{.8in} \frac{1}{2}$}
\newsavebox{\sqcc}
\savebox{\sqcc}{
\put(26,26){\line(1,0){94}}
\put(20,20){\line(1,0){100}}
\put(26,26){\line(0,1){94}}
\put(20,20){\line(0,1){100}}
\put(20,20){\circle*{10}}
\put(120,23){\circle*{10}}
\put(23,120){\circle*{10}}}
\put(350,0){\usebox{\sqcc}}
\end{picture}
\end{figure}
%---------------------------------------------------------------
\vspace{6truecm}
\begin{center}
FIG. 2
\end{center}
%
%----------------------------------------------------------------------fig 3
\newpage
\begin{figure}[htb]
\setlength{\unitlength}{.3mm}
\begin{picture}(400,200)(30,0)
\newsavebox{\fg}
\savebox{\fg}{
\put(0,0){\line(0,1){200}}
\put(0,100){\circle*{2}}
\put(-15,100){$x_1$}}
\put(200,0){\usebox{\fg}}
\newsavebox{\fghi}
\savebox{\fghi}{
\put(0,0){\line(0,1){200}}
\put(0,100){\circle*{2}}
\put(-15,100){$x_3$}}
\put(400,0){\usebox{\fghi}}
\newsavebox{\fgh}
\savebox{\fgh}{
\put(0,0){\line(1,0){50}}
\put(0,50){\line(1,0){50}}
\put(0,0){\line(0,1){50}}
\put(50,0){\line(0,1){50}}
\put(0,0){\circle*{2}}
\put(0,-15){$x_2$}
}
\put(275,100){\usebox{\fgh}}
\end{picture}
\end{figure}
\vspace{6truecm}
\begin{center}
FIG. 3
\end{center}
%
%---------------------------------------------------------------fig 2
\newpage
\begin{figure}[htb]
\setlength{\unitlength}{.3mm}
\begin{picture}(400,200)(30,0)
\newsavebox{\rst}
\savebox{\rst}{
\put(0,0){\line(1,0){30}}
\put(0,30){\line(1,0){30}}
\put(0,0){\line(0,1){30}}
\put(30,0){\line(0,1){30}}
\put(-18,-5){$+$}
\put(-18,30){$+$}
\put(37,-5){$+$}
\put(37,30){$+$}
\put(10,-15){1}
\put(10,35){3}
\put(-15,10){4}
\put(37,10){2}
}
\newsavebox{\rstu}
\savebox{\rstu}{
\put(0,0){\line(1,0){30}}
\put(0,30){\line(1,0){30}}
\put(0,0){\line(0,1){30}}
\put(30,0){\line(0,1){30}}
\put(-18,-5){$-$}
\put(-18,30){$-$}
\put(37,-5){$-$}
\put(37,30){$-$}
\put(10,-15){1}
\put(10,35){3}
\put(-15,10){4}
\put(37,10){2}
}
\newsavebox{\rstuv}
\savebox{\rstuv}{
\put(0,0){\line(1,0){30}}
\put(0,30){\line(1,0){30}}
\put(0,0){\line(0,1){30}}
\put(30,0){\line(0,1){30}}
\put(-18,-5){$+$}
\put(-18,30){$+$}
\put(37,-5){$-$}
\put(37,30){$-$}
\put(10,-15){1}
\put(10,35){3}
\put(-15,10){4}
\put(37,10){2}
}
\newsavebox{\rstuvw}
\savebox{\rstuvw}{
\put(0,0){\line(1,0){30}}
\put(0,30){\line(1,0){30}}
\put(0,0){\line(0,1){30}}
\put(30,0){\line(0,1){30}}
\put(-18,-5){$-$}
\put(-18,30){$-$}
\put(37,-5){$+$}
\put(37,30){$+$}
\put(10,-15){1}
\put(10,35){3}
\put(-15,10){4}
\put(37,10){2}
}
\put(50,25){\usebox{\rstu}}
\put(125,35){$-\sin(\theta_1 +\theta_2 -\theta_3 -\theta_4)$}
\put(50,125){\usebox{\rst}}
\put(125,135){$\sin(\theta_1 +\theta_2 -\theta_3 -\theta_4)$}
\put(350,25){\usebox{\rstuvw}}
\put(425,35){$-\sin(\chi_1 -\theta_2 -\chi_3 -\theta_4)$}
\put(350,125){\usebox{\rstuv}}
\put(425,135){$\sin(\chi_1 -\theta_2 -\chi_3 -\theta_4)$}
\end{picture}
\end{figure}
%---------------------------------------------------------------
\vspace{6truecm}
\begin{center}
FIG. 4
\end{center}
%

%----------------------------------------------------------------------fig 3
\newpage
\begin{figure}[htb]
\setlength{\unitlength}{.3mm}
\begin{picture}(400,200)(30,0)
\newsavebox{\sqa}
\savebox{\sqa}{
\put(0,0){\line(0,1){200}}
\put(0,100){\circle*{2}}
\put(-15,100){$x$}}
\put(100,0){\usebox{\sqa}}
\newsavebox{\sqb}
\savebox{\sqb}{
\put(0,0){\line(1,0){25}}
\put(0,25){\line(1,0){25}}
\put(0,0){\line(0,1){25}}
\put(25,0){\line(0,1){25}}
\put(0,-15){$y$}
\put(0,0){\circle*{6}}}
\put(150,100){\usebox{\sqb}}
\put(50,175){(a)}
\newsavebox{\tqa}
\savebox{\tqa}{
\put(0,0){\line(0,1){200}}
\put(0,100){\circle*{2}}
\put(-15,100){$x$}}
\put(290,0){\usebox{\tqa}}
\newsavebox{\tqb}
\savebox{\tqb}{
\put(0,0){\line(1,0){25}}
\put(0,25){\line(1,0){25}}
\put(0,0){\line(0,1){25}}
\put(25,0){\line(0,1){25}}
\put(0,-15){$y$}
\put(0,0){\circle*{2}}}
\put(340,100){\usebox{\tqb}}
\put(240,175){(b)}
\newsavebox{\sqqc}
\savebox{\sqqc}{
\put(0,0){\line(0,1){100}}
\put(0,103){\line(0,1){97}}
\put(0,100){\line(1,0){50}}
\put(0,103){\line(1,0){50}}
\put(-15,100){$x$}}
\put(480,0){\usebox{\sqqc}}
\newsavebox{\sqd}
\savebox{\sqd}{
\put(0,0){\line(1,0){25}}
\put(0,25){\line(1,0){25}}
\put(0,3){\line(0,1){22}}
\put(25,0){\line(0,1){25}}
\put(0,-15){$y$}}
\put(530,100){\usebox{\sqd}}
\put(430,175){(c)}
\end{picture}
\end{figure}
\vspace{6truecm}
\begin{center}
FIG. 5
\end{center}
%
%
%
%----------------------------------------------------------------------
\newpage
\begin{samepage}

\begin{figure}[htp]
\vspace{-1.0cm}
\begin{minipage}{0.49\textwidth}
\epsfig{file=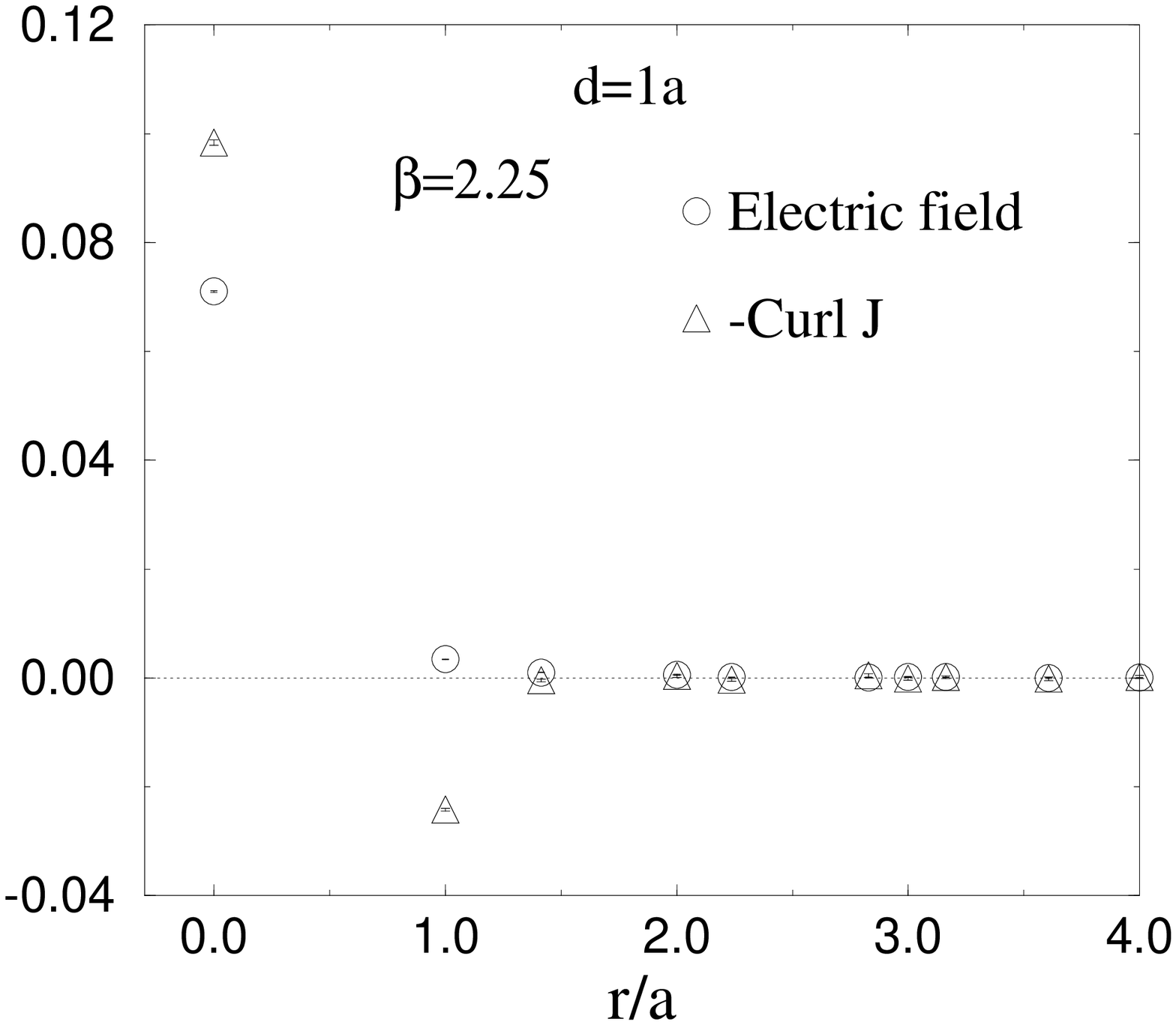,width=6.8truecm,angle=-90}
\end{minipage}
\begin{minipage}{0.46\textwidth}
\epsfig{file=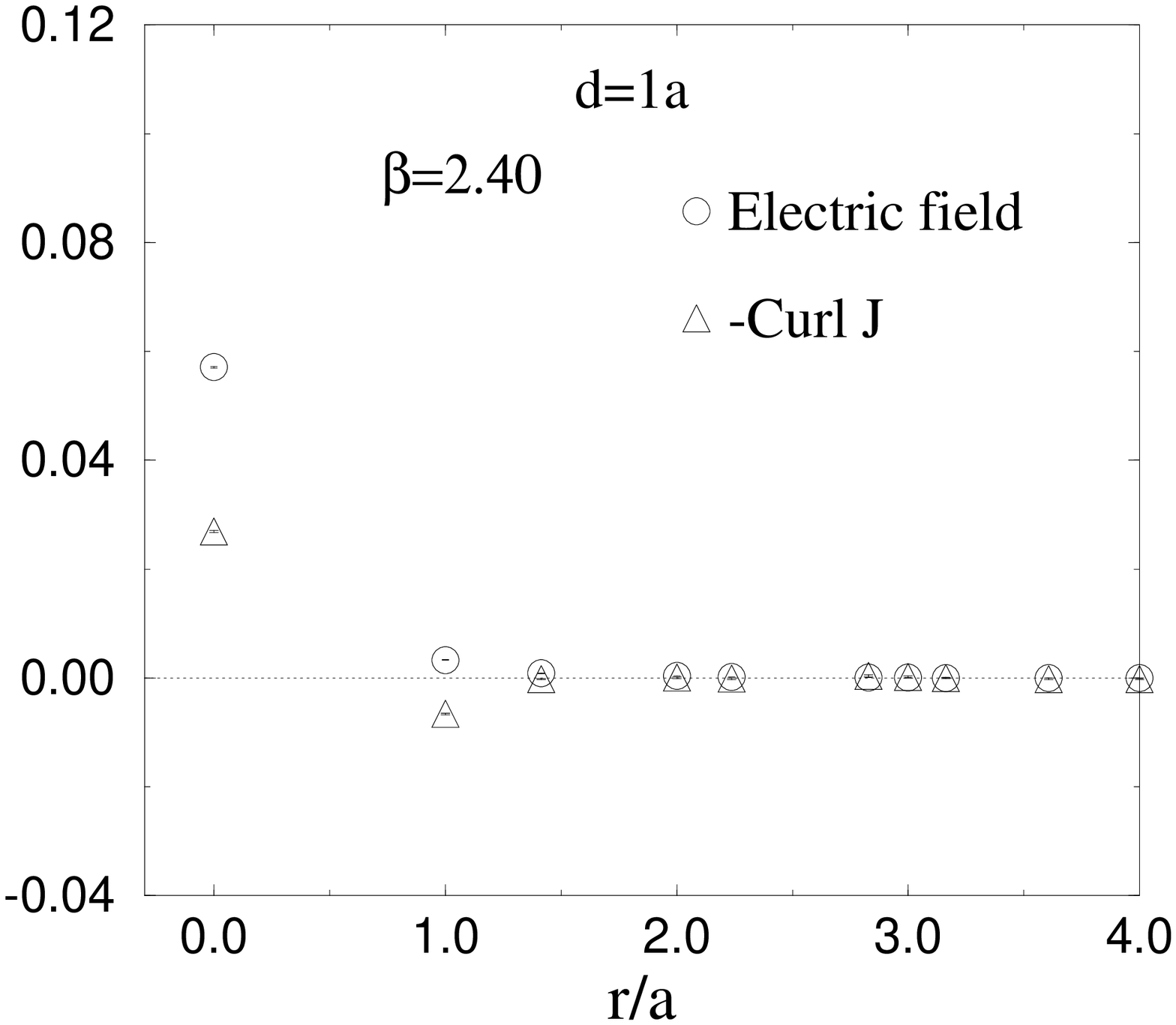,width=6.8truecm,angle=-90}
\end{minipage}
\begin{minipage}{0.49\textwidth}
\epsfig{file=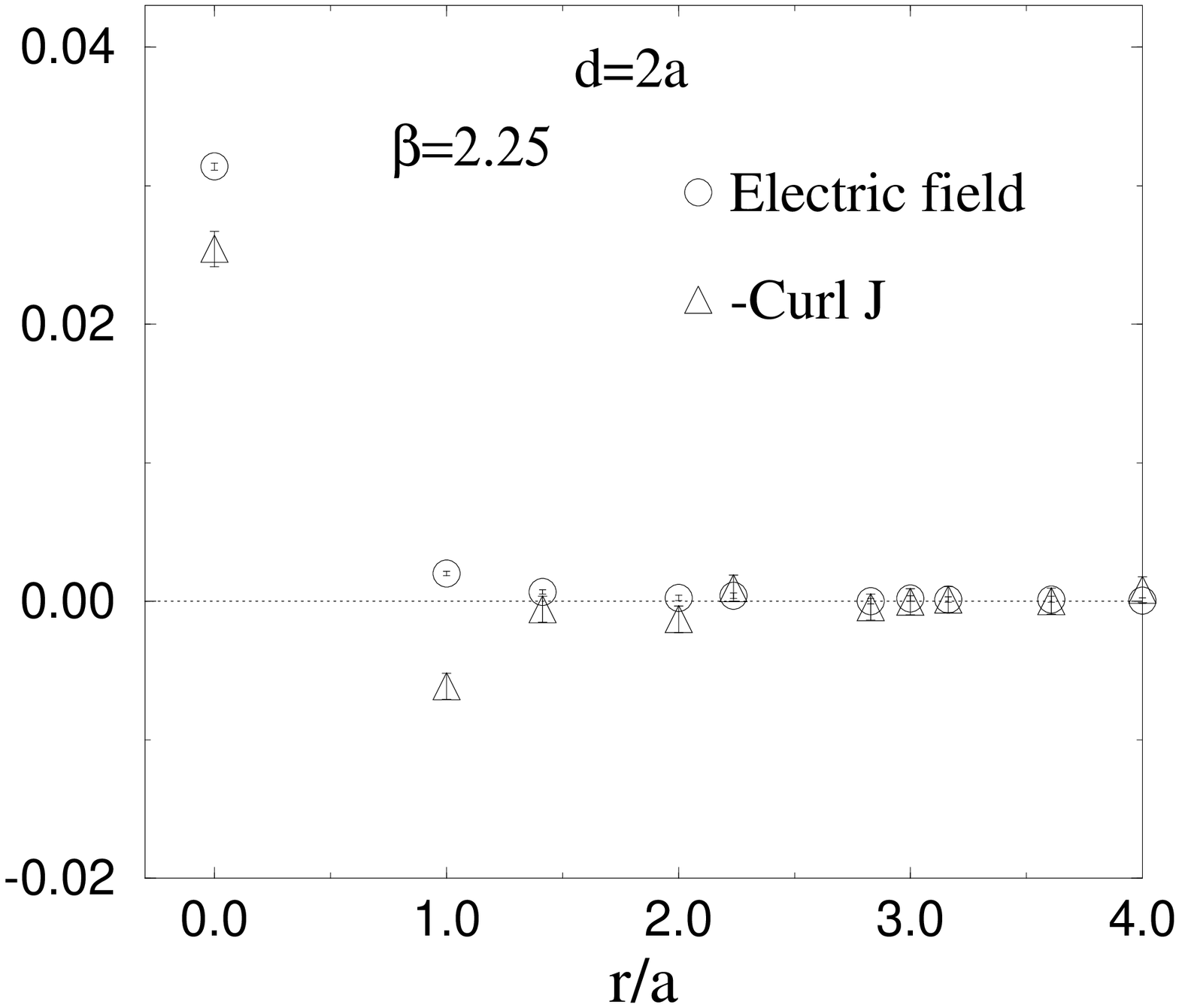,width=6.8truecm,angle=-90}
\end{minipage}
\begin{minipage}{0.49\textwidth}
\epsfig{file=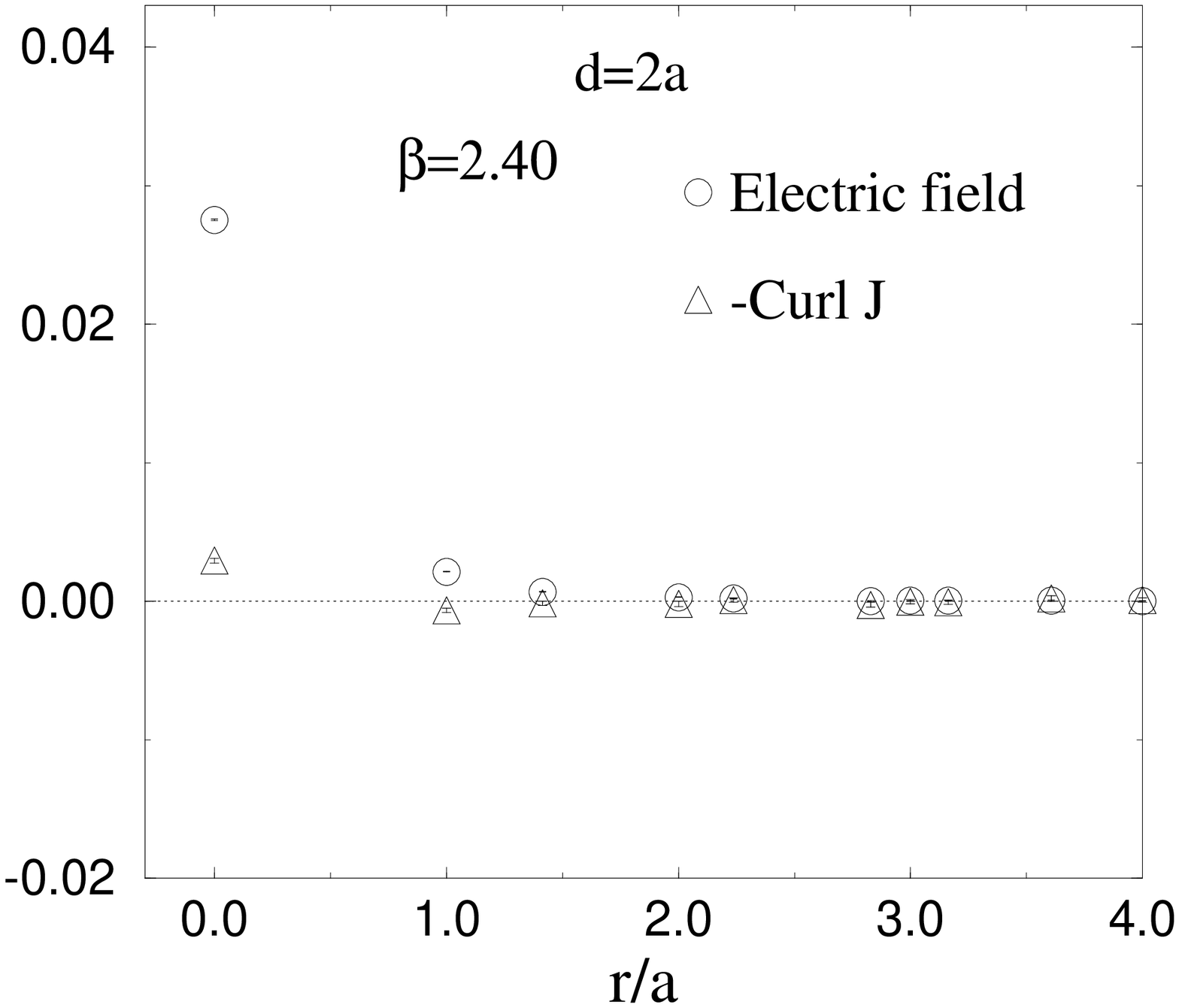,width=6.8truecm,angle=-90}
\end{minipage}
\begin{minipage}{0.49\textwidth}
\epsfig{file=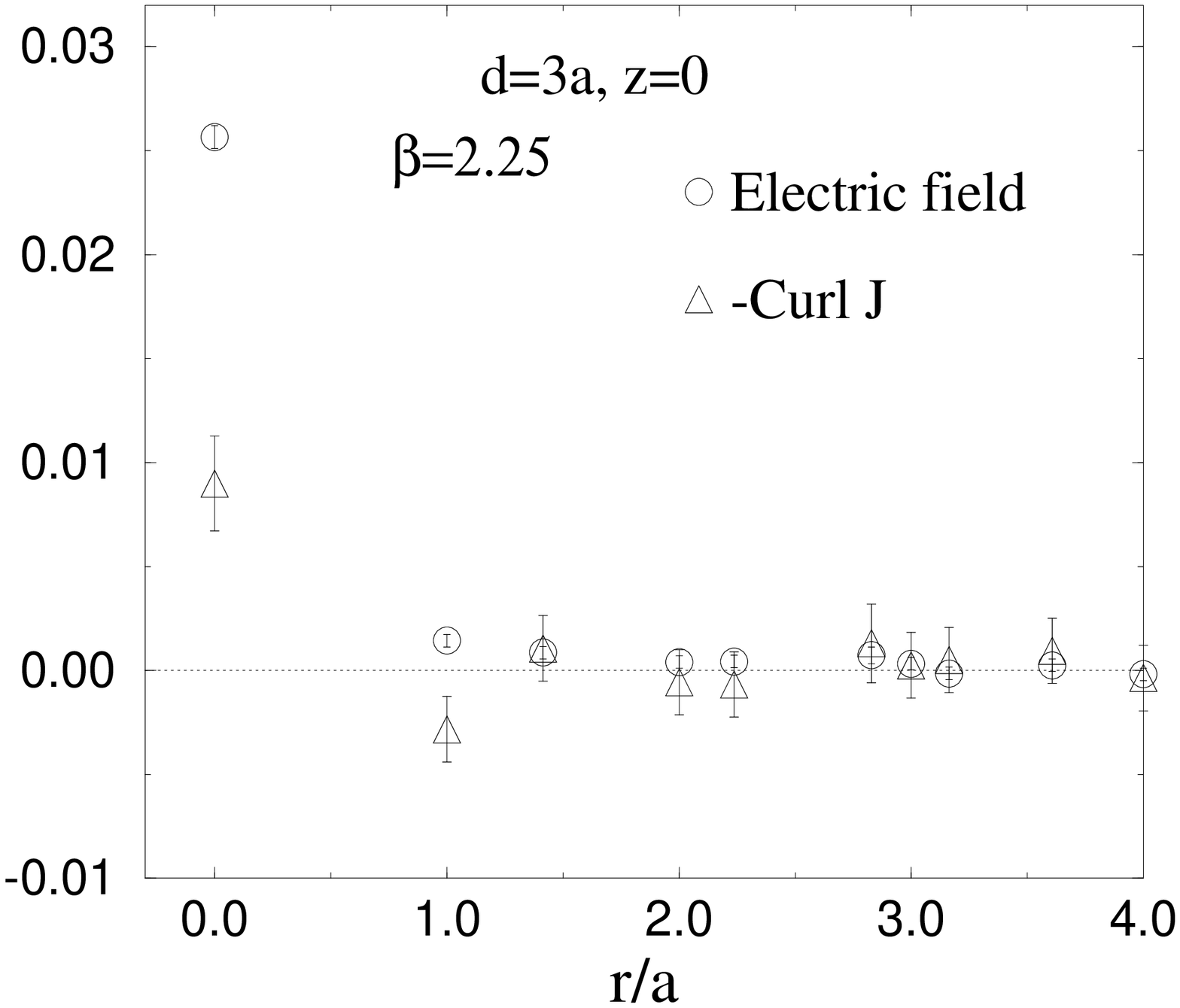,width=6.8truecm,angle=-90}
\end{minipage}
\begin{minipage}{0.49\textwidth}
\epsfig{file=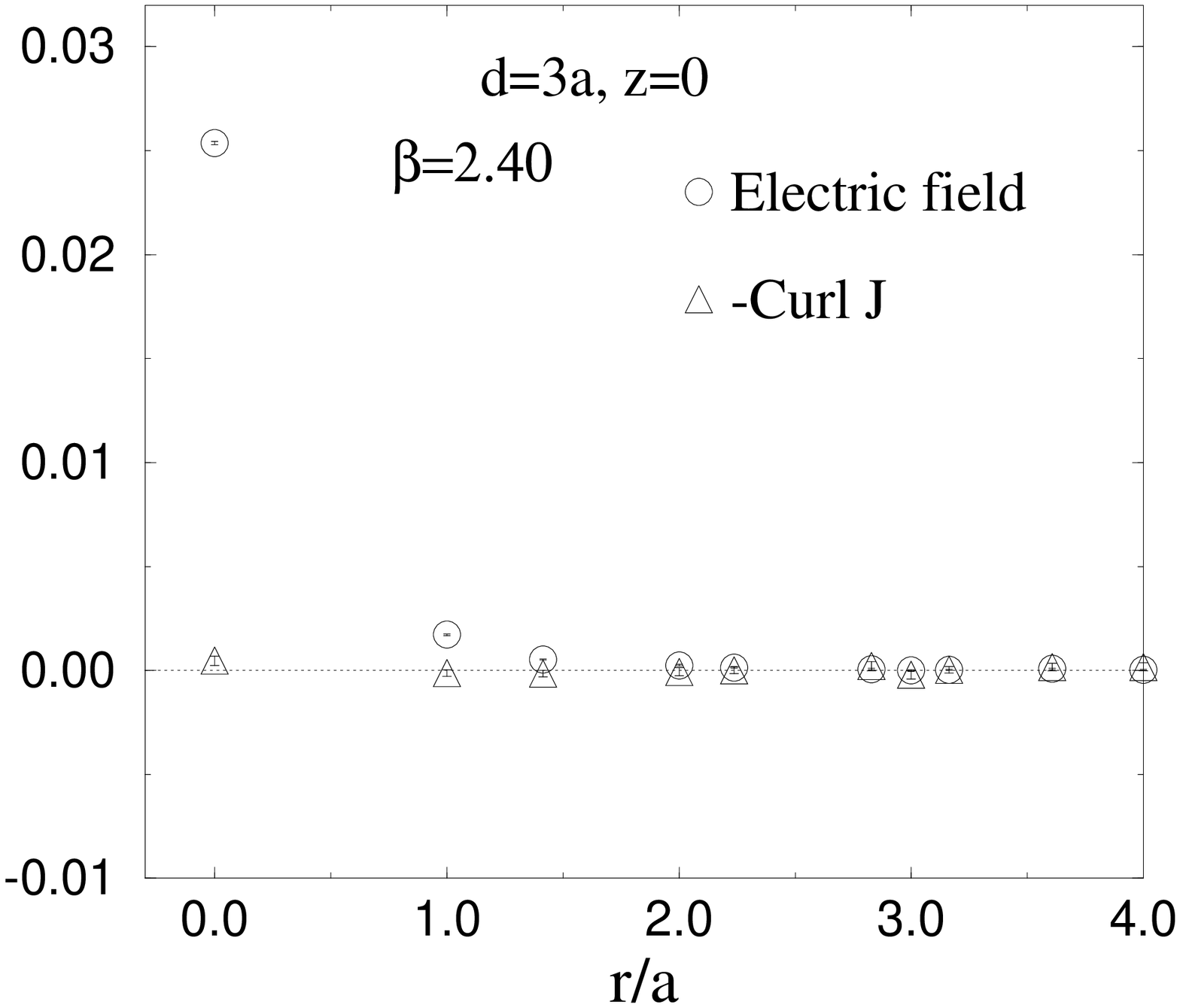,width=6.8truecm,angle=-90}
\end{minipage}
\end{figure}
\nopagebreak
\vspace{0.4truecm}
\begin{center}
FIG. 6
\end{center}
\end{samepage}
\newpage
\begin{figure}[htpb]
\begin{minipage}{0.46\textwidth}
\epsfig{file=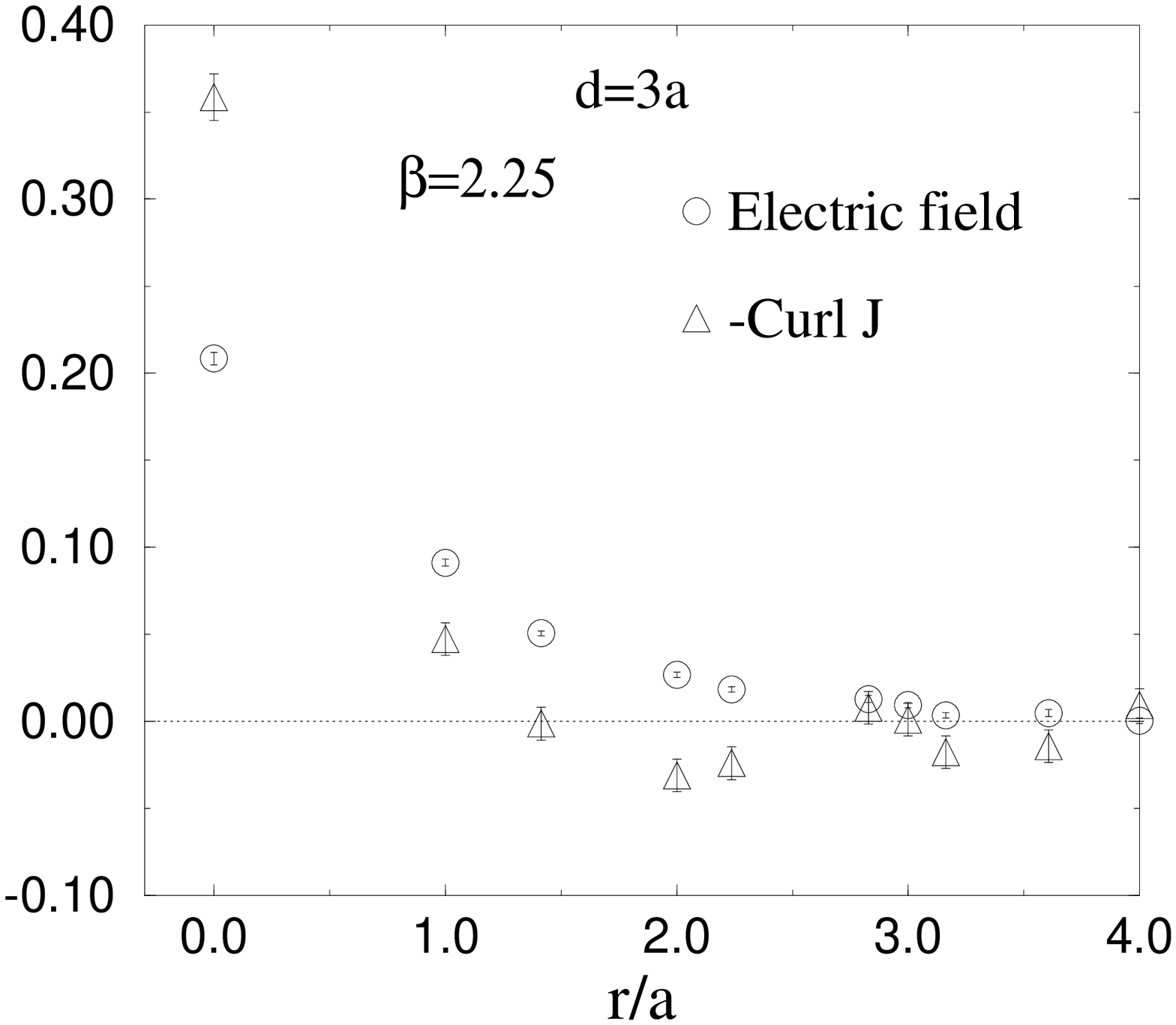,width=6.8truecm,angle=-90}
\end{minipage}
\begin{minipage}{0.46\textwidth}
\epsfig{file=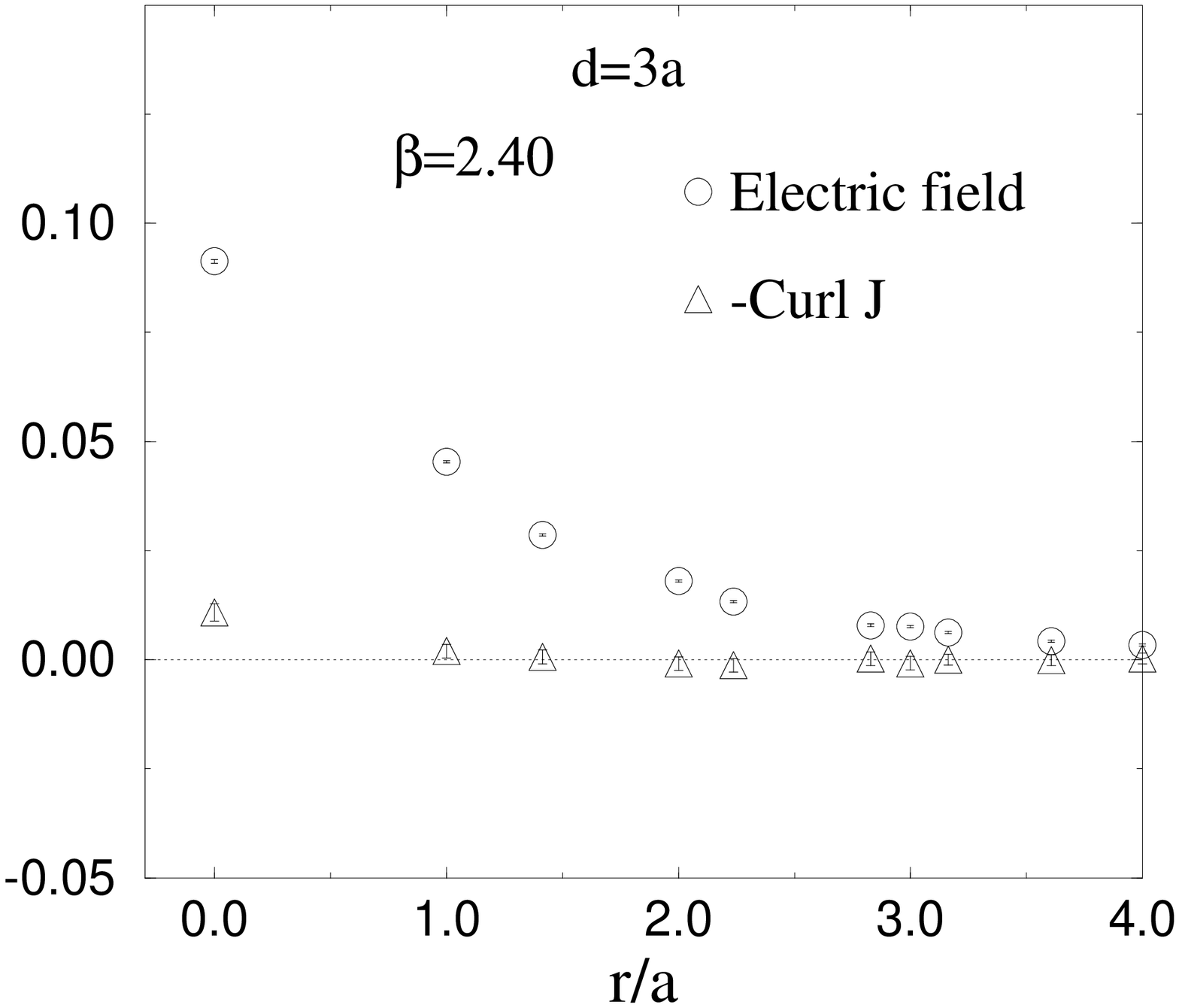,width=6.8truecm,angle=-90}
\end{minipage}
\end{figure}
\vspace{6truecm}
\begin{center}
FIG. 7
\end{center}
\newpage
\begin{figure}[htpb]\hspace{1.8truecm}
\epsfig{file=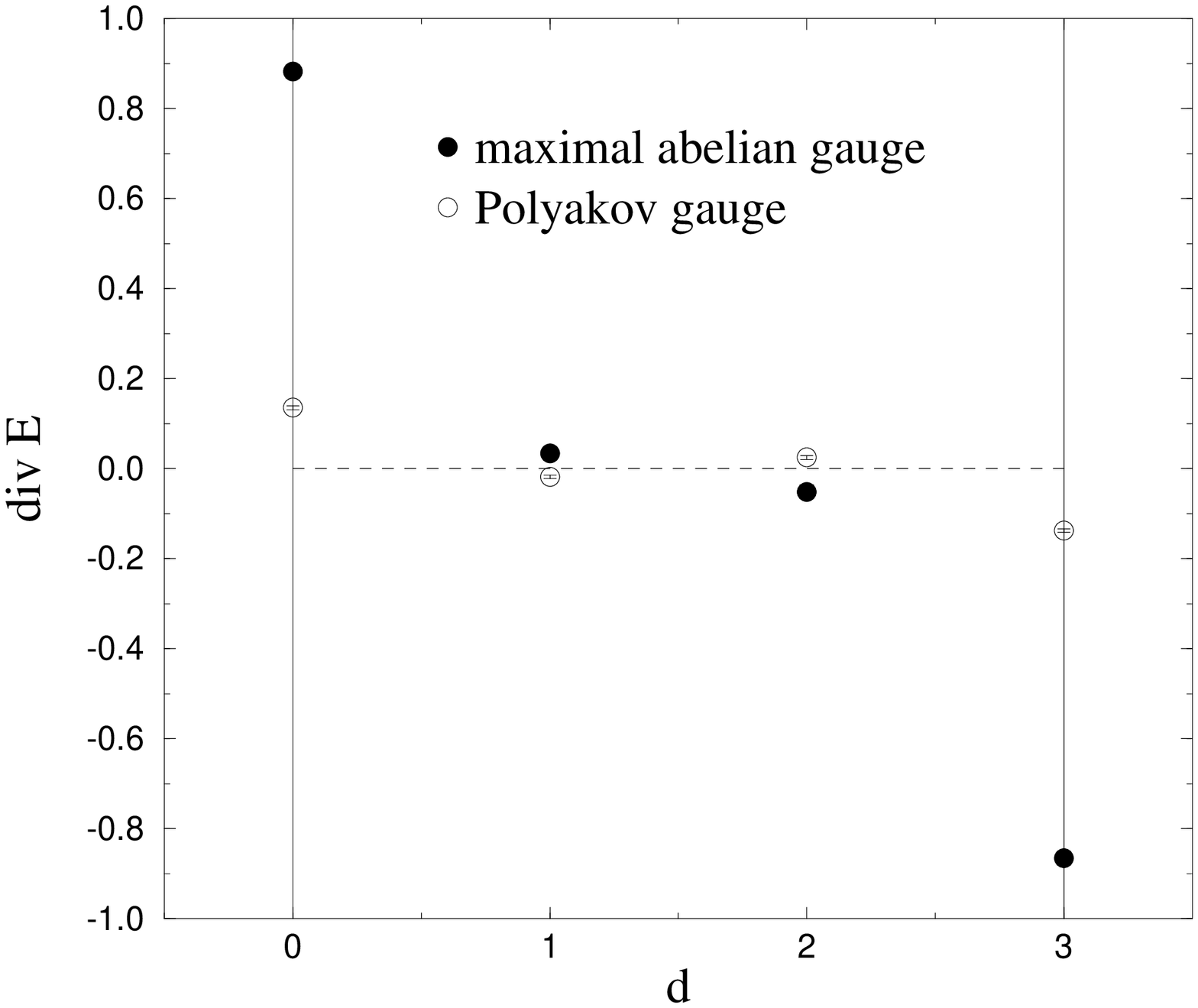,width=10truecm,angle=-90}
\end{figure}
\vspace{6truecm}
\begin{center}
FIG. 8
\end{center}


\begin{thebibliography}{99}
%
\bibitem{weinberg}  S. Weinberg, 
\underline{The Quantum Theory of Fields} Vol II,Cambridge University
Press, Cambridge 1996, See chapter 21.6 for a thorough treatment of
this viewpoint.
%
\bibitem{hay} For a review and further references 
see e.g. R.W. Haymaker, {\it ``Dual Abrikosov 
vortices in U(1) and SU(2) lattice gauge theories''}, Proceedings of 
the international School of Physics ``Enrico Fermi'', Course 
CXXX, A. Di Giacomo and D. Diakonov (Eds.) IOS Press, 
Amsterdam 1996, hep-lat 9510035.
%
\bibitem{suzuki} For a review of monopoles and confinement, see e.g.
T. Suzuki, Nucl.Phys. {\bf B} Proc.  Suppl. {\bf 30}, 176 (1993).
%
\bibitem{gl} V. L. Ginzburg and L. D. Landau, Zh. Ekxperim, i Theor. Fiz,
{\bf 20}, 1064 (1950).
%
\bibitem{abrikosov} A. A. Abrikosov, Sov. Phys. JETP {\bf 32}, 1442 
(1957).
%
\bibitem{tinkham} M. Tinkham, Introduction to Superconductivity, (McGraw-Hill,
 New York, 1975)
%
\bibitem{fm} J. Fr\"{o}lich and P. A. Marchetti, Euro. Phys. Lett. {\bf 2},
933 (1986); Commun. Math. Phys. {\bf 112}, 343 (1987);
{\bf 116}, 127 (1988);
{\bf 121}, 177 (1989); Lett. Math. Phys {\bf 16}, 347 (1988).
%
\bibitem{pw} L. Polley and U.-J. Wiese, Nucl. Phys. {\bf B356}, 621 (1991).
%
\bibitem{ppw} M. I. Polikarpov, L. Polley and U.-J. Weise, Phys. Lett. {\bf B253}, 212 (1991).
%
\bibitem{giacomo2} L. Del Debbio, A. Di Giacomo and G. Paffuti, Phys. Lett.
{\bf B349}, 513, (1995); Nucl. Phys B (Proc. Suppl.) {\bf 42}, 231 (1995);
%
\bibitem{shb} V. Singh, R.W. Haymaker, and
 D.A. Browne, Phys. Rev. {\bf D47},
1715 (1993).
%
\bibitem{hsbw} R.W. Haymaker, V. Singh, D. Browne and J. Wosiek; Proc. of 
Workshop on QCD Vacuum Structure and its applications, The American 
University of Paris, June 1-5, 1992, p184, World Scientific 1993.
%
\bibitem{thooft1} G. 't Hooft, Nucl. Phys. {\bf B 190}, 455 (1981).
%
\bibitem{cpv2} M.N. Chernodub, M.I. Polikarpov and A.I. Veselov, 
Phys. Lett. {\bf B 342}, 303 (1995).
%
\bibitem{ksww}
A.S. Kronfeld, G. Schierholz 
and U.J. Wiese, 
Nucl. Phys. {\bf B 293}, 461 (1987);
%
\bibitem{klsw} A.S. Kronfeld, M.L. Laursen, G. Schierholz and U.J. Wiese, 
Phys. Lett. {\bf B 198}, 516 (1987);   T. Suzuki and I. Yotsuyanagi,
Phys. Rev. {\bf D 42}, 4257 (1990).
%
\bibitem{ddpp} L. Del Debbio, A. Di Giacomo, G. Paffuti and P. Pieri, 
Phys. Lett. {\bf B 355}, 255 (1995).
%
\bibitem{sbh} V. Singh, D.A. Browne, and R.W. Haymaker, Nucl. Phys. B
(Proc. Suppl.) {\bf 30}, 658 (1993);
 Phys. Lett. {\bf B306}, 115 (1993).
%
\bibitem{mes} Y. Matsubara, S. Ejiri, and T. Suzuki, Nucl. Phys. {\bf B34}
(Proc. Suppl.) 176 (1994).
%
\bibitem{ph} Y. Peng and R.W. Haymaker,  Phys. Rev.  {\bf D 52} 3030,
(1995).
%
\bibitem{cc}P. Cea and L. Cosmai, Nucl. Phys. {\bf B} Proc. Suppl. {\bf 30},
572 (1993); Phys. Rev. {\bf D 52}, 5152 (1995).
%
\bibitem{ss} H. Shiba and T. Suzuki, Phys. Lett. {\bf B 333}, 461 (1994).
%
\bibitem{sw} J. Stack and R. Wensley, Nucl. Phys. {\bf D 22}, 597 (1992).
%
\bibitem{snw} J. Stack, S. Neiman and R. Wensley, Phys. Rev. {\bf D 50},
3399 (1994)
%
\bibitem{bbmps} G. S. Bali, V. Bornyakov, M. M\"{u}ller-Preussker and
K. Schilling, Phys. Rev. {\bf D 54} 2863 (1996).
%
\bibitem{cpv1} M.N. Chernodub, M.I. Polikarpov and A.I. Veselov, 
Nucl. Phys. Proc. Suppl. {\bf 49},  307, (1996).
%
\bibitem{nbekms} N. Nakamura, V. Bornyakov, 
S. Ejiri, S. Kitahara, Y. Matsubara and T. Suzuki, Lattice-96 
conference abstract, St. Louis, June 1996 e-print 9608004.
%
\bibitem{sfz} P. Skala, M. Faber and M. Zach, 
Ahrenshoop Symp. 301-306, 1995  
hep-lat/9603009.
%
\bibitem{gribov} V.N. Gribov, Nucl. Phys. {\bf B 139}, 1 (1978).
%
\bibitem{gg} H. Georgi and S.L. Glashow, 
Phys. Rev. Lett. {\bf 28}, 1494 (1972).
%
\bibitem{thooft2} G. 't Hooft, Nucl. Phys. {\bf B 79}, 276 (1974).
%
\bibitem{zfks} M. Zach, M. Faber, W. Kainz and P. Skala, Phys. Lett. 
{\bf B 358}, 325 (1995).
%
\bibitem{simoy} T. Suzuki, S. Ilyar, Y. Matsubara, T. Okude and K. Yotsuji,
Phys. Lett. {\bf B 347}, 375 (1995); {\it ERRATUM-ibid.} {\bf B 351}, 
603 (1995).
%
\end{thebibliography}
\end{document}